\documentclass[twocolumn]{aastex631}

\newcommand{\mz}{\textit{m/z}}
\newcommand{\cmu}{cm$^{-1}$ }

\usepackage{multirow}

\shorttitle{AASTeX v6.3.1 Sample article}
\shortauthors{Piacentino et al.}
\graphicspath{{./}{figures/}}

\begin{document}

\title{Characterization of monosubstituted benzene ices}

\author[0000-0001-6947-7411]{Elettra L. Piacentino}
\affiliation{Harvard-Smithsonian Center for Astrophysics, 60 Garden Street, Cambridge, MA 02138, USA }

\author[0009-0006-4436-9848]{Aurelia Balkanski}
\affiliation{Harvard-Smithsonian Center for Astrophysics, 60 Garden Street, Cambridge, MA 02138, USA }

\author[0000-0003-2761-4312]{Mahesh Rajappan}
\affiliation{Harvard-Smithsonian Center for Astrophysics, 60 Garden Street, Cambridge, MA 02138, USA }

\author[0000-0001-8798-1347]{Karin I. {\"O}berg}
\affiliation{Harvard-Smithsonian Center for Astrophysics, 60 Garden Street, Cambridge, MA 02138, USA }

\begin{abstract}

Aromatic structures are fundamental for key biological molecules such as RNA and metabolites and the abundances of aromatic molecules on young planets are therefore of high interest. Recent detections of benzonitrile and other aromatic compounds in interstellar clouds and comets have revealed a rich aromatic astrochemistry. In the cold phases of star and planet formation, most of these aromatic molecules are likely to reside in icy grain mantles, where they could be observed through IR spectroscopy.
We present laboratory IR spectra of benzene and four monosubstituted benzene molecules -- toluene, phenol, benzonitrile and benzaldehyde -- to determine their IR ice absorbances in undiluted aromatic ices, and in mixtures with water and CO.
We also characterize the aromatic ice desorption rates, and extract binding energies and respective pre-exponential factors using temperature programmed desorption experiments. We use these to predict at which protostellar and protoplanetary disk temperatures these molecules sublimate into the gas-phase. We find that benzene and mono-substituted benzene derivatives are low-volatility with binding energies in the 5220-8390 K (43-70 kJ/mol) range, which suggests that most of the chemistry of benzene and of functionalized aromatic molecules is to be expected to occur in the ice phase during star and planet formation.

\end{abstract}

\section{Introduction}

Aromatic molecules are one of the most important classes of molecules for biochemical functionality. Particularly when containing heteroatoms and functional groups, aromatic molecule can easily form strong interactions with each other by overlap of the electrons clouds on the rings ($\pi$ stacking). In macromolecules, this $\pi$ stacking capability leads to the formation of stable molecular configurations that express different chemical properties and specific reactivity. This propriety becomes particularly relevant in molecules such as DNA, RNA and enzymes, where the presence of aromatic moieties in specific sections of the molecular chains is crucial for their correct biochemical functionalities. 
Benzene, constituted of only carbon and hydrogen atoms, is the prototypical aromatic hydrocarbon. The functionalization of benzene to form monosubstituted benzene is one of the first possible steps to build chemical complexity from aromatic precursors. 

Large aromatic molecules, so-called polycyclic aromatic hydrocarbons (PAHs), have been known to exist in a range of astrophysical environments for many decades  (see e.g. \citet{leger1984identification,allamandola1989interstellar,tielens1997circumstellar,tielens2013molecular}), and they have therefore been subject of a large number of laboratory studies including, \citet{gudipati2003ApJ...596L.195G,gudipati2006ApJ...638..286G,bouwman2010A&A...511A..33B,cook2015photochemistry,Zeichner2023Sci...382.1411Z}. Benzene was first observed in a protoplanetary nebula CRL 618 \citep{cernicharo2001infrared}, but it is only recently that small aromatics such as benzonitrile, indene and 1-cyanonaphtalene, have been detected in the cold star-forming regions of the interstellar medium (ISM) \citep{mcguire20182018,mcguire2021detection,burkhardt2021discovery}

The presence of these relatively complex aromatic molecules in the ISM, and in remnants of the cold solar nebula (i.e. comets), suggests that star and planet formation is characterized by a rich aromatic astrochemistry \citep{kaiser2015ARPC...66...43K,oberg2016arXiv160903112O}, which includes a large range of substituted benzene molecules. Being of relatively high molecular weight ($>$ 78 amu), aromatic molecules are expected to have desorption temperature above that of water ($\sim$ 160 K in a laboratory setting) and consequently much of their chemistry should take place on icy grains, where they will readily condense out at the temperatures characteristic of molecular clouds and in protoplanetary disks beyond the water snowline. A few laboratory studies on the functionalization of benzene in ice analogs exist in the literature showing that these reactions can proceed at low temperature for the formation of a variety of substituted benzene molecules. Carboxylic acids, for example, have been show to form upon electron irradiation of an ice mixture of benzene and CO$_2$ \citep{mcmurtry2016formation}. The formation of benzonitrile can be achieved in the ice via electron irradiation of mixtures of benzene and acetonitrile \citep{maksyutenko2022formation}. Heterocyclic species could also be formed in the ice. \citet{materese2015n} detected N- and O-heterocycle molecules in the residues derived from the UV irradiation of ice mixtures composed of benzene in either a water or ammonia matrix.

With the advent of the new James Webb Space Telescope (JWST), the detection of rare IR-active molecules has become easier to achieve \citep{mclure2023NatAs...7..431M}. As of now, JWST data have already revealed the presence of gas-phase benzene in the inner disk of the J160532 M dwarf star \citep{tabone2023rich}. JWST may also have the sensitivity to detect the presence of aromatic molecules in the icy grains in interstellar clouds, protostellar envelopes, and protoplanetary disks, and in icy solar system objects \citep{mclure2023NatAs...7..431M,yang2022ApJ...941L..13Y,kospa2023ApJ...945L...7K,emery2023arXiv230915230E}. Initial observations of such objects have already revealed several new ice constituents \citep{2021A&A...651A..95T}, and careful comparison between observed spectra and spectroscopic ice data on larger organic molecules is likely to yield additional detections. 

Characterizing the observational icy aromatic inventory requires precise laboratory infrared spectra of small aromatic molecules in interstellar and solar system ice analogs (i.e. in mixture with more abundant molecules). Existing experimental studies have provided detailed spectra of benzene and pyridine \citep{hudson2022infrared}, but most monosubstituted benzene molecules remain poorly characterized. A second aspect of molecular characterization concerns their binding energies, which directly determines both under which condition a molecule is in the gas vs. solid phase, and indirectly how readily the molecule can diffuse in an ice and therefore partake in chemical reactions. Only few studies are available on the sublimation behavior of benzene and toluene (for binding energies measurements see :\citet{salter_tpd_2018,thrower_thermal_2009,zacharia_interlayer_2004}, for details about the vapor pressure and the condensation behavior of benzene see \citet{HudsonBenVP2022PSJ.....3..120H} and \citet{Dubois2021PSJ.....2..121D}) and, to our knowledge none are available for other small aromatic molecules. The characterization of small aromatic molecules in ice analog is not only of aid for comparison with observational spectra, but it also provides constrains needed to explore aromatic ice chemistry in the laboratory.

With the aim of gathering spectral data and information of the the characteristics desorbtion kinetics of small aromatic molecules, we survey benzene, toluene, phenol, benzonitrile and benzaldehyde ices under ultra-high vacuum (UHV) conditions. 
We determine their IR ice signatures in both undiluted (one-component) (\S\ref{sec:IR-10k}) and mixed ices (\S\ref{sec:IR mix}). The mixed ice spectra are most relevant for comparison with astrophysical sources, while the undiluted ices spectra are key for the interpretation of the mixed ice spectroscopy.
We also characterize the desorption behavior of undiluted ices  of each of the four benzene derivatives determining their desorption temperature range (\S\ref{sec:DES}) and binding energies (\S\ref{sec:BE}). In \S\ref{sec:AI} we use the new laboratory data to put upper limits on the aromatic ice reservoir using recent JWST ice spectra of a prestellar core, and estimate the fractionation of these small aromatic molecules between ice and gas in a simple protoplanetary disk model.

\section{Experimental Methods}\label{sec:met}

\subsection{Experimental Set-up}
We experimentally characterize mono substituted derivatives of benzene using two apparatuses, SPACECAT (Surface Processing Apparatus for Chemical Experimentation to Constrain Astrophysical Theories) and SPACEKITTEN (Surface Processing Apparatus for Chemical Experimentation - Kinetics of Ice Transformation in Thermal ENvironments). The full capabilities of both instruments are described in detail elsewhere \citep{martin2020formation,Simon2023ApJ...955....5S}.

In brief, both instruments consist of an Ultra-High Vacuum (UHV) chamber that is pumped down to reach $~$10$^{-9}$, 10$^{-10}$ torr respectively for SPACECAT and SPACEKITTEN. 
A helium cryocooled CsI substrate is mounted at the center of the spherical chambers. Ices are grown from pure or mixed gas-phase samples which are condensed onto the CsI substrate using a gas dosing assembly consisting of a staineless steel tube ( $\diameter$ 4.8 mm) placed at 17.8 mm from the substrate surface. The main difference between the two set-ups is the that SPACECAT gas line can handle corrosive gases, while the SPACEKITTEN gas line cannot. We therefore used SPACECAT exclusively for benzonitrile experiments, and both set-ups, dependent on availability, for the other molecules.
Once deposited on the substrate, and throughout the duration of the experiments, the ices are monitored using a Fourier Transform Infrared Spectrometer (FTIR)
(Bruker Vertex 70V and Bruker Vertex 70) respectively for SPACECAT and SPACEKITTEN) with a liquid N$_2$ cooled mercury cadmium telluride (MCT) detector in transmission mode.

In both set-ups the substrate temperature can be controlled between 12 and 300 K using a temperature controller (LakeShore 335) that has an accuracy of $\pm$2 K  and uncertainty of 0.1 K. 
During the temperature programmed desorption (TPD) experiments, the substrate is heated at a constant rate of 2 K/min causing sublimation of the sample. The desorbing molecules are detected using a quadrupole mass spectrometer (QMS) (Pfeiffer QMG 220M1 and Pfeiffer QMG 220M2,
PrismaPlus Compact respectively for SPACECAT and SPACEKITTEN) up to a temperature of 280 K.

\subsection{Reagents and sample preparation}
The chemicals used in this work are the following:
C$_6$H$_6$ (MilliporeSigma, 99.8$\%$), C$_6$H$_5$OH (MilliporeSigma, 99$\%$), C$_6$H$_5$CN (MilliporeSigma, 99.9 $\%$), C$_6$H$_5$CH$_3$ (MilliporeSigma, 99.8$\%$), C$_6$H$_5$CHO, (MilliporeSigma ,99$\%$), CO (MilliporeSigma 99.9$\%$), and purified water.

Benzene, toluene and benzonitrile were transferred directly into a resealable pre-evacuated flask and attached to the instrument gas line. In the case of phenol a small quantity of solid was transferred into a clean flask which was then sealed and evacuated to $\sim$10$^{-4}$ Torr. To avoid contamination of the gas line with solid particulates during pumping the flask is equipped with a stainless steel Tee-Type particulate filter (440 micron pore size). Despite the low vapor pressure of phenol no significant variation in the sample preparation procedure was necessary as a suitable amount of phenol could be dosed into the gas line directly by opening the flask. Benzaldehyde is known to oxidize to benzoic acid when exposed to air \citep{2014NatCo...5.3332S}. Therefore, we transfered benzaldehyde to the flask under nitrogen atmosphere inside a glove bag. All samples were further purified by several freeze-thaw-pump cycles using liquid nitrogen. 
For most of our experiments we used a ice deposition rate in the range of 0.14-4.3 x 10$^{16}$ molecules/min. Previous studies have shown that for small molecules amorphous ice is produced at very low deposition rates while higher rate will induce crystallization of the ice.  Amorphous CH$_4$ ice for example is formed at deposition rates of ~ 1.5 x 10$^{17}$ molecules/min \citep{GerakinesCO2015ApJ...805L..20G} while amorphous CO is produced at deposition rates below  3.3 x 10$^{14}$ molecules/min \citep{Gerakines2023MNRAS.522.3145G}. Comparison of our benzene spectra with the one reported by \citet{hudson2022infrared} for amorphous benzene shows that no crystalline benzene features are present in our spectra confirming the amorphous nature of the ices. In some of the benzonitrile experiments the deposition rate was slightly higher (1-3 x 10$^{17}$ molecules/min) but spectral comparison with benzonitrile ices deposited at rates (2.4 x 10$^{16}$ molecules/min) comparable with the one used by \citep{hudson2022infrared}, confirm that no crystallization is induced by the higher deposition rates in the benzonitrile ices.

\subsection{Estimation of the IR band strengths}
The IR band strengths of the majority of the investigated aromatic ices have not been reported in the literature. Accurate band strength measurement can be obtained in the laboratory using laser interferometry \citep{hudson2022infrared} which is currently not available on any of our experimental set-ups.
To obtain approximate ice coverages, which are needed to extract pre-exponential factors and binding energies from our desorption experiments, we therefore follow \citet{1997ApJ...476..932B} to estimate band strengths from ice mixtures with a species with a known band strength. The procedures and resulting band strength approximations are reported in Appendix \ref{appir}, together with an analysis of the uncertainty of this method. The estimated band strength uncertainty of 50 $\%$ for all ices, except for benzene, is then used when calculating uncertainties in our desorption rates and binding energies.
We emphasize that the values reported in Appendix \ref{appir} are provisional estimates and they should ideally be replaced by more precise values using laser interferometry \citep{hudson2022infrared}.

\subsection{Extracting desorption energies and pre-exponential factors} \label{met:be}
We determined the binding energies in the multilayer regime of these molecules using the Polyani-Wigner formalism.
While the astrophysically most relevant binding energy for a molecule is often considered the one to a water surface, in the case of molecules that are less refractory than water we expect that water ice would already have desorbed. Furthermore for refractory or "stickier" molecules the specific nature of the surface appears to matter less than for hypervolatiles \citep{salter2021using} making the multi-layer binding energy relevant for estimating the astrophysical behavior. To model the desorbing behavior of the molecules  we use the Polyani-Wigner equation applied to the TPD curves:

\begin{equation} \label{eqn1}
-\frac{d\theta}{dT} = \frac{\nu}{\beta} \theta^{n}e^{-E_{ \rm Des}/T}
\end{equation}

where d$\theta$/dT is the desorption rate, $\nu$ is the pre-exponential factor, $\beta$ is the heating rate, $\theta$ is the ice coverage in monolayers, $n$ is the kinetic order and in the multilayer regime is 0, $E_{des}$ is the binding energy of the molecule, and T is the temperature of the ice.

The preexponential factor $\nu$ represent the attempt frequency of vibration of a molecule trying to escape the ice matrix and it can be estimated using different methods. Traditionally, the value of of the attempted frequency $\nu$ has been estimated in the harmonic oscillator approximation using:

\begin{equation} \label{eqn2}
    \nu_{har} = \sqrt{\frac{2N_{ \rm s}E_{ \rm Des}}{\pi^2 \mu m_{ \rm H}}}
\end{equation}

where N$_{ \rm s}$ is the binding site density (generally fixed at $10^{15}$ cm$^{-2}$), 
$\mu$$m_{ \rm H}$ is the mass of the molecule, and  $E_{\rm Des}$ is the binding energy of the molecule that, can be calculated iteratively from experimental data (e.g. using the \textit{lmfit} python package) \citep{bergnerHCN2022ApJ...933..206B,Acharyya07,Noble12,Fayolle16}.
While this method yields a good approximation in the case of small molecules, it provides a poor fit for bigger molecules where the variation of the rotational energy contribution between the adsorbed and gas-phase molecules cannot be neglected \citep{MinissaleTST2022ESC.....6..597M,FerreroTST2022MNRAS.516.2586F}. 

An alternative way to theoretically estimate $\nu$ while accounting for the variation in the rotational partition function of the desorbing molecule, is to use the transition state theory (TST) model described most recently in \citet{MinissaleTST2022ESC.....6..597M}. 
In this method the rotational and translational partition functions of the molecules are calculated using:

\begin{equation} \label{eqn3}
    Q_{ \rm Trs}^{\ddagger}=\frac{B}{{(\frac{h}{\sqrt{2m\pi K_{ \rm b} T_{ \rm Des}}}})^2}
\end{equation}

and

\begin{equation} \label{eqn4}
    Q_{ \rm Rot}^{\ddagger} = (\frac{\sqrt{\pi}}{\sigma h^3})  (8 \pi^2 k_{ \rm b} T_{\rm Des})^\frac{3}{2}   \sqrt{I_{x}I_{y}I_{z}}
\end{equation}

$\nu_{ \rm TST}$ can then be derived using:

\begin{equation} \label{eqn5}
    \nu_{ \rm TST} = \frac{k_{ \rm b} T_{\rm Des}}{h}Q_{\rm Trs}^{\ddagger} Q_{\rm Rot}^{\ddagger}
\end{equation}

Where $Q_{ \rm Trs}^{\ddagger}$ $Q_{\rm Rot}^{\ddagger}$ are the 2D translations and the 3D rotational partition function for a molecule in the transition state from adsorbed to free,{\textit{B} is the number of binding sites per unit area, $m$ is the mass of the molecules, $I_{\rm x}I_{\rm y}$ and $I_{\rm z}$ are the principal moment of inertia, $\sigma$ is the symmetry factor and $T_{\rm Des}$ is the temperature at which the molecule desorption peaks.

The attempt frequency value can also be determined directly from TPD experiments by fitting the TPD curve using a Polyani-Wigner model which simultaneously fits for both $\nu$ and $E_{\rm Des}$ \citep{bergnerHCN2022ApJ...933..206B}.  This method generally requires fitting multiple experiments to yield accurate results.

In this work we calculate the binding energies and attempt frequencies/pre-exponential factors using all three methods and evaluate their respective usefulness for the specific case of small aromatic molecules.

\section{IR Spectroscopy of monosubstituted benzene ices} \label{sec:IR}
This section reports the identification of the main IR band in  the undiluted aromatic ice spectra at 10 K (\S\ref{sec:IR-10k}), the spectroscopic effects associated with embedding the aromatic molecules in CO or H$_2$O ice (\S\ref{sec:IR mix}), and  spectral changes during ice warm-up (\S\ref{Sec:IR_VS_T}). 
For clarity, are listed in Table \ref{Table_big}. The molecular geometries of the aromatic molecules are shown in Fig. \ref{fig:molecules} where the atoms are numbered to ease the identification of specific atoms in the text.

\startlongtable
\tablewidth{0.5\columnwidth}
\begin{deluxetable*}{cccc|cccc}
\tabletypesize{\scriptsize}
\tablecolumns{8}
\tablecaption{IR band position and relative integrated intensity} \label{Table_big}
\tablehead{ \multicolumn{4}{c|}{Band Position (\cmu)}&\multicolumn{4}{c}{Rel. Integrated Intensity}}

\startdata
 10 K &  T$_{tr}$ & in CO & in H$_2$O & 10 K & T$_{tr}$ & in CO & in H$_2$O \\
\hline
\multicolumn{8}{c}{Benzene (T$_{tr}$= 60 K)}\\
\hline
676   &  680  & 688   & 685  & \textbf{1.00} &  \textbf{1.00}   & \textbf{1.00}   & \textbf{1.00} \\
-     &  706  &  -    & 685  & -    &  0.04   & -      &  -   \\
1036  &  1033 & 1038  & 1036 & 0.13 &  0.20   & 0.017  & 0.05 \\
-     &   -   & 1100  & 1100 &  -   &  -   & 0.26   & 0.11  \\
1477  & 1477  & 1482  & 1480 & 0.28 & 0.44    & 0.40   & 0.21 \\
\hline
\multicolumn{8}{c}{Toluene (T$_{tr}$= 140 K) }\\
\hline
695   & 695   & 698   & 699   & 0.33 & 0.42 & 0.25 & 0.27 \\
731   & 731   &  730    & 738   & \textbf{1.00} & \textbf{1.00} & \textbf{1.00} & \textbf{1.00} \\
  -   & 1027  &  -    & -     & -    & 0.04 & -    & -    \\
1030  & 1032  & 1031  & 1032  & 0.10 & 0.06 & 0.09 & 0.03 \\
1042  & 1041  & 1042  & 1045  & 0.05 & 0.02 & 0.08 &  -   \\
1082  & 1082  & 1083  & 1085  & 0.10 & 0.09 & 0.29 & 0.06 \\
 -    & -     & 1100  & 1099  & -    & -    & 0.24 & 0.24 \\
1467  & 1462  & 1469  & 1469  & 0.25 & 0.16 & 0.35 & 0.18 \\
1495  & 1495  & 1497  & 1498  & 0.31 & 0.26 & 0.5  & 0.36 \\
1605  & 1604  & 1607  & 1607  & 0.10 & 0.08 & 0.16 & 0.09 \\
\hline
\multicolumn{8}{c}{Phenol (T$_{tr}$= 200 K)}\\
\hline
690   & 691   & 691   &   -   & 0.62 & 0.58 & 0.23 & -   \\
755   & 754   & 759   &   -   & 1.51 & 1.94 & 0.70 & 0.33   \\
810   & 812   & 812   &   -   & 0.39 & 0.15 & 0.20 & 0.10  \\
825   & 821   & 826   & 825   & 0.26 & 0.63 & 0.26 & 0.44   \\
889   & 889   & -     &   -   & 0.18 & 0.10 & 0.03 & -   \\
1027  & 1024  & 1027  &  -    & 0.11 & 0.09 & 0.02 & -    \\
1072  & 1072  & 1072  & 1100  & 0.21 & 0.20 & 0.06 & 1.33    \\
  -   &  -    & 1198  &   -   & -    & -    & 0.53 & -   \\
1235  & 1239  & 1233  & 1245  & 4.18 & 5.42 & 0.27 & 2.22   \\
  -   &  -    & 1260  &   -   & -    & -    & 0.43 & -  \\
1478  & 1476  & 1474  &   -   & 1.80 & 2.0 & 0.56  & -    \\
1503  & 1500  & 1503  & 1509  & \textbf{1.00} & \textbf{1.00} & \textbf{1.00} & \textbf{1.00}    \\
1598  & 1596  & 1598  &  -    & 1.96 & 1.92 & 1.2 & -   \\
  -   &  -    & 3575  &  -    &   -   &   -  & 1.7  & - \\
  -   &  -    & 3590  &  -    &   -   &   -  & 1.25 & - \\
\hline
\multicolumn{8}{c}{Benzonitrile (T$_{tr}$= 170 K)}\\
\hline
689   & 686   & 691   & 688   & 0.59  & 0.51 & 0.48 & 0.55 \\
763   & 773   & 766   & 761   & \textbf{1.00}  & \textbf{1.00} & \textbf{1.00} & \textbf{1.00} \\
930   & 937   & 933   &   -   & 0.11  & 0.05 & 0.08 & -    \\
1000  & 998   & 998   &   -   & 0.02  & 0.01 & 0.01 & -    \\
-     & 1008  &  -    &   -   &  -    & 0.01 & -    & -    \\
1028  & 1025  & 1028  & 1028  & 0.07  & 0.02 & 0.10 & 0.08 \\
1074  & 1076  & 1074  & 1074  & 0.06  & 0.04 & 0.07 & 0.06 \\
1161  & 1165  & 1163  & 1163  & 0.03  & 0.04 & 0.01 & 0.02 \\
1179  & 1179  & 1181  & 1181  & 0.04  & 0.11 & 0.06 & 0.06 \\
1290  & 1290  & 1290  & 1290  & 0.06  & 0.03 & 0.07 & 0.28 \\
1447  & 1450  & 1449  & 1449  & 0.25  & 0.17 & 0.30 & 0.34 \\
1491  & 1491  & 1493  & 1491  & 0.27  & 0.17 & 0.42 & 0.34 \\
1599  &  -    & 1601  & 1601  & 0.05  & -    & 0.01 & 0.11 \\
2230  & 2227  & 2233  & 2239  & 0.68  & 0.69 & 0.74 & 1.40 \\
\hline
\multicolumn{8}{c}{Benzaldehyde (T$_{tr}$= N.A.)}\\
\hline
650   &  -    & 650   & 652   & 0.10 & - & 0.08 & 0.12  \\
689   &  -    & 691   & 688   & 0.12 & - & 0.08 & 0.10  \\
750   &  -    & 753   & 750   & 0.27 & - & 0.18 & 0.02  \\
827   &  -    & 829   & 833   & 0.13 & - & 0.09  & 0.10 \\
1170  &  -    & 1168  & 1172  & 0.07 & - & 0.09  & 0.09 \\
1207  &  -    & 1205  & 1213  & 0.29 & - & 0.27  & 0.31 \\
1313  &  -    & 1313  & 1315  & 0.06 & - & 0.05  & 0.04 \\
1394  &  -    & 1396  & 1400  & 0.05 & - & 0.05  & 0.05 \\
1456  &  -    & 1458  & 1458  & 0.07 & - & 0.05  & 0.08 \\
1584  &  -    & 1586  & 1586  & 0.07 & - & 0.09  & 0.11 \\
1599  &  -    & 1599  & 1601  & 0.10 & - & 0.09  & 0.11 \\
1700  &  -    & 1707  & 1690  & \textbf{1.00} & - & \textbf{1.00}  & \textbf{1.00} \\
\enddata
\vspace{-0.1cm}
\tablenotetext{*}{T$_{tr}$ is the temperature at which the morphology of the ice change as per IR spectra variation. Integrated intensity values are normalized to the most intense aromatic peak.}
\end{deluxetable*}

\subsection{IR spectra of undiluted aromatic ices at 10 K}\label{sec:IR-10k}
The IR spectra of each of the undiluted aromatic ices acquired at 10 K are shown in Fig. \ref{fig:IR_nor} where the strongest spectral features are marked with their wavenumbers. When available, the IR bands are assigned to specific vibrational mode by comparison with ice spectra reported in the literature. When not available, gas-phase or liquid phase spectra are used instead.

\begin{figure*}[thb!]
  \centering
  \includegraphics[width=\textwidth]{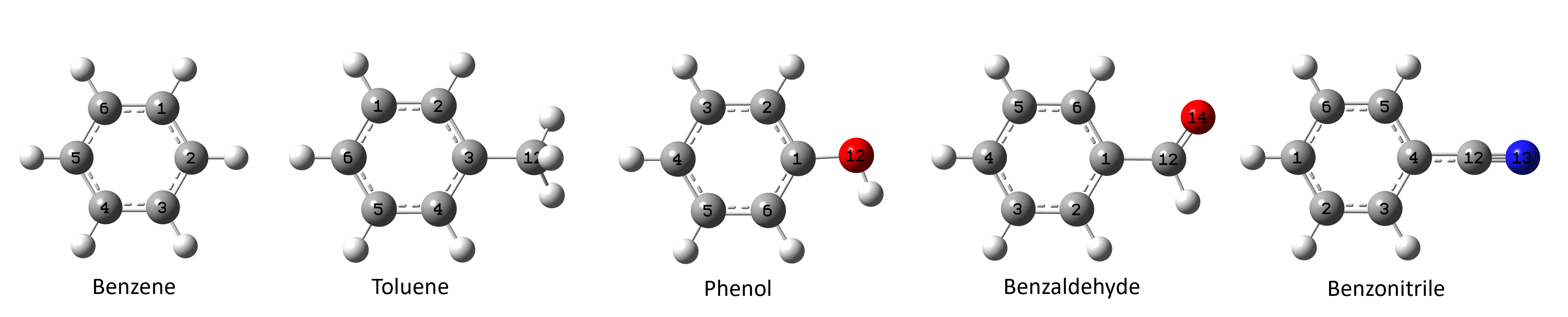}
 \caption{Molecular geometries of the mono-substituted benzene molecules. Atoms are numbered for clarity.}
 \label{fig:molecules}
\end{figure*}

The 10 K spectra of benzene can be directly compared to the spectra of benzene ice reported by \cite{hudson2022infrared} and \cite{mouzay2021experimental}. Consistent with the spectral assignments from \citet{mouzay2021experimental}, our benzene ice spectra shows the out-of-plane (oop) libration of the H atoms near 676 \cmu  as well as the peak at 1477 \cmu due to the C=C stretch on the ring. Additionally, the in-plane wagging of the hydrogen atoms is visible near 1036 \cmu, while the weak cluster near 3000 \cmu is due to the stretch of C-H bonds. 

\begin{figure*}[thb!]
  \centering
  \includegraphics[width=\textwidth]{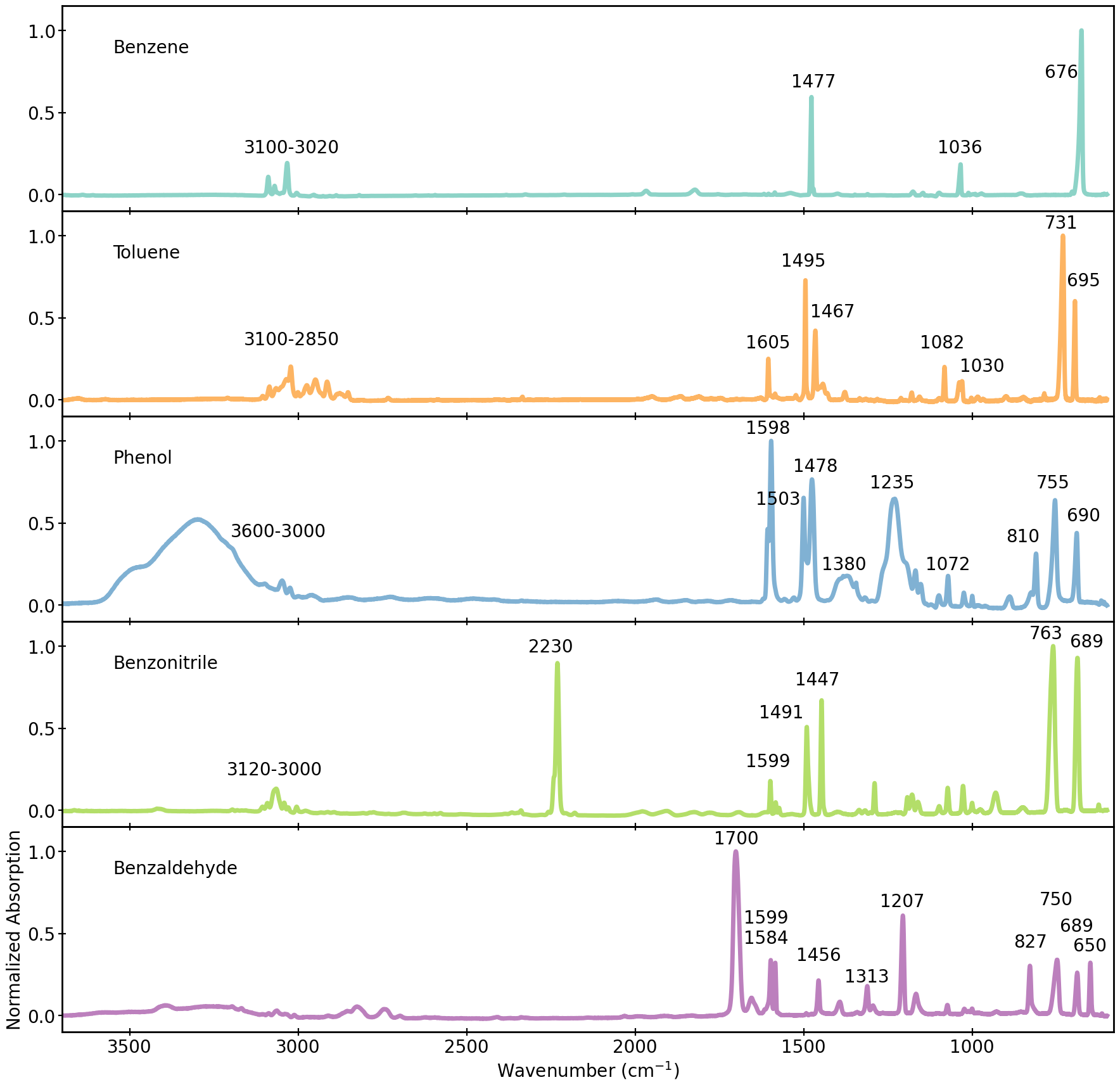}
 \caption{Infrared spectra of the undiluted ices of aromatic molecules at 10 K. The position of the IR features is labeled.}
 \label{fig:IR_nor}
 
\end{figure*}

The ice spectra of toluene at 25 K has been previously measured by \citet{salter_tpd_2018} and we follow their assignments. The spectra of toluene present its most prominent features at 731 and 695 \cmu. Similarly to the benzene spectra, these features are due to the oop vibrational mode of the hydrogen but, given the presence of the methyl group, the mode is decoupled into symmetric (731 \cmu) and asymmetric oop (695 \cmu) modes. In the range 1000-1100 \cmu toluene presents a feature at about 1030 \cmu, caused by the bending mode of the methyl hydrogens, and a second feature at 1082 \cmu due to a combination mode dominated by the in-plane rocking of the hydrogen atoms on the ring. In the 1400-1700 \cmu range we see three peaks, where the two lowest wavenumber are due to asymmetric and symmetric C=C stretch while the highest wavenumber band is characteristic of the C-C stretch between the methyl group and the C3 on the ring.

In the absence of IR spectra of ice phase phenol in the literature we assign the IR peaks detected in our experiment by comparison with the gas-phase spectra reported by \citet{lampert1997molecular}.  The O-H stretch of phenol is visible in the 3600-3000 \cmu range, while combination modes are seen at 1380 and 1235 \cmu \citep{lampert1997molecular}. The three ring stretching modes are fairly strong and are visible in the 1478-1598 \cmu range. Similarly to the spectra of toluene, the hydrogen in-plane bending modes are seen in the 1028-1072 \cmu range while the three out-of-plane vibrational modes are visible below 810 \cmu.

The IR spectra of benzonitrile has been previously reported by \citet{green1976vibrational} for the liquid and vapour phase of the molecule. The vibrational band reported by \citet{green1976vibrational} are close to the peaks observed in the ice spectra allowing for assignment by direct comparison. The C=N stretch is the dominant feature of the benzonitrile IR spectra at 2230 \cmu.
The symmetric, asymmetric, and combination C=C stretches are visible at 1600 and 1491 and 1447 \cmu respectively while the two in-plane and  three oop bending modes of the benzonitrile hydrogens are seen below 1100 \cmu as observed for other molecules.

We assign the IR feature of benzaldehyde by similarity with its gas phase spectra reported by \citet{zwarich1971assignment}.The spectra of benzaldehyde is dominated by the C=O stretch at 1700 \cmu and followed in intensity by the 1207 \cmu  C1-C(OH) stretching feature. Similar to the other mono-substituted benzene ices the three peaks in the 1400-1600 \cmu range are due to ring C=C stretches and the C-H bending modes are visible below 850 \cmu.

It is noticeable that the presence of a ring substituent cause a peak splitting in the oop and in-plane C-H vibrational regions. This effect has been reported before in both liquid and solid phase spectra of monosubstituted benzene and it correlates with the disruption of the degeneracy of the C-H vibrational modes due to the reduced molecular symmetry \citep{cannon1951infra}. The magnitude of the frequency shift depends on the nature of the substituent with bigger shifts as the substituent is more electron withdrawing \citep{kross1956infrared}

\subsection{IR spectra of mono-substituted benzene species in CO and H$_2$O ice} \label{sec:IR mix}

In Fig. \ref{fig:IR_CO} and \ref{fig:IR_H2O_sub} we show the spectra of benzene, toluene, phenol, benzaldehyde and benzonitrile diluted in either CO or H$_2$O with a ratio 1:10. In the text we describe the differences and similarities of the diluted ice spectra in comparison with the spectra of the aromatic molecules ice discussed in the previous section. 

\subsubsection{Aromatic molecules embedded in a CO matrix} \label{subsubco} 
The 10 K spectra of the aromatic:CO mixtures in a 1:10 ratio are generally dominated by the CO feature at 2139 \cmu, however all the major features of the aromatic molecule remain visible and the shapes of the peaks do not change significantly compared to the spectra of undiluted ices. To focus the attention on the monosubstituted benzene molecules, the spectra are plotted on a compressed vertical axis that cuts out most of the CO feature. Common to all the molecules surveyed we find that the aromatic IR peaks are generally shifted toward higher wavenumbers (+1-12 \cmu or 0.002-0.33 $\mu$m) in the presence of a CO matrix. In addition we find that hydrogen oop modes are consistently more shifted than ring related vibrational modes as might be expected with a matrix-mediated perturbation of the ice structure (Table \ref{Table_big}). Band narrowing is also appreciable in the lower wavenumber spectral range (See Appendix \ref{appB}, Fig. \ref{fig:nar}) across all the molecules studied. 

\begin{figure*}[thb!]
  \centering
  \includegraphics[width=\textwidth]{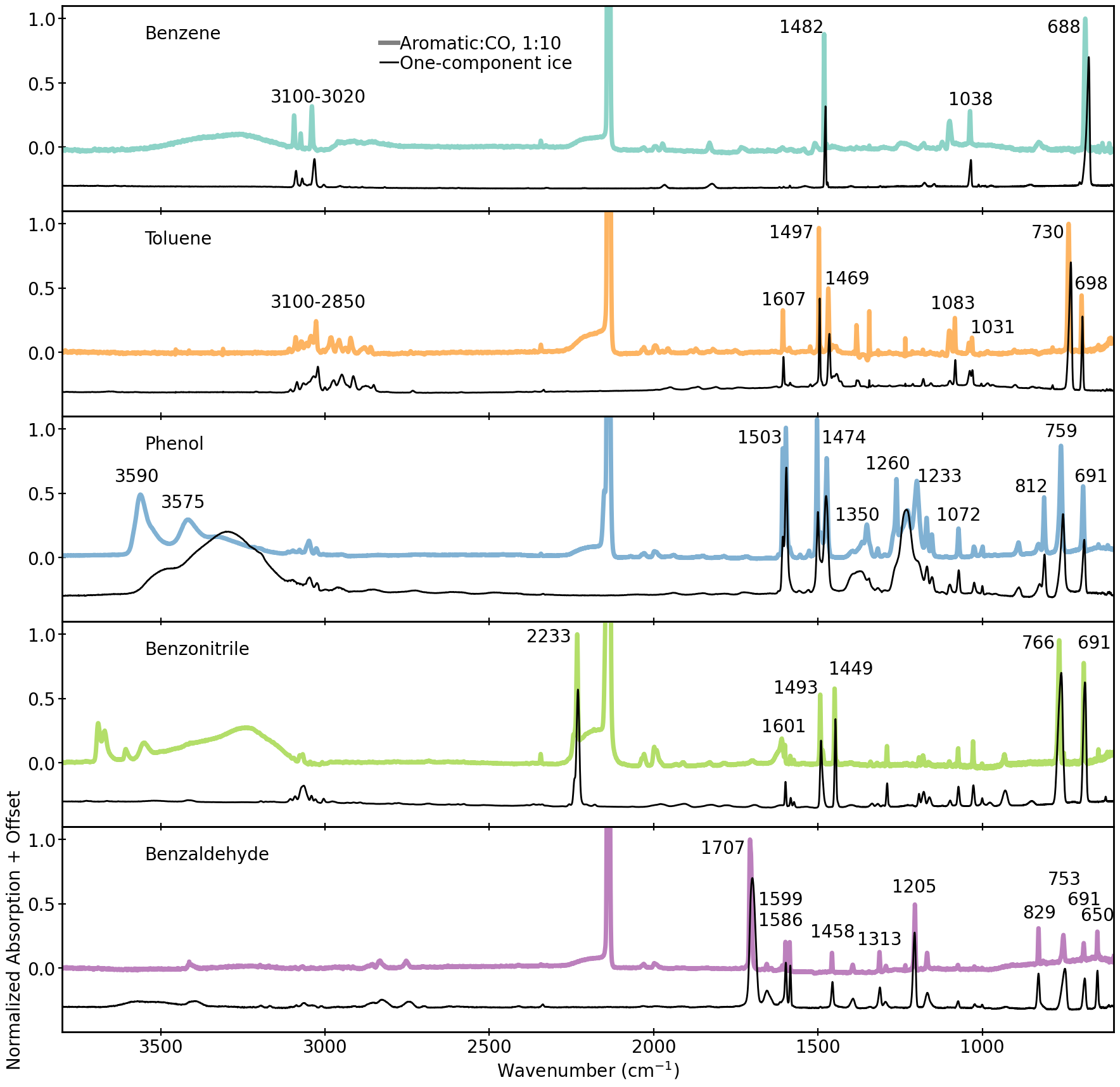}
 \caption{Normalized spectra of ices condensed from a  aromatic:CO 1:10 mixture at 10 K. The black trace shows the 10 K ice spectra of the aromatic molecules ices. The band at 3300 \cmu in the benzonitrile spectra is due to a small water contamination.}
 \label{fig:IR_CO}
\end{figure*}

Beyond the peak shifts and feature narrowing, benzene and phenol presents a couple of more profound changes in their CO-mixture spectra. 
More precisely, in the 1:10 benzene:CO (Fig. \ref{fig:IR_CO}) we see the appearance of a feature at 1100 \cmu that is not present in the spectra of undiluted benzene ice. We speculate that this feature is due to the formation of a $\pi$ interaction between benzene and CO \citep{nowak1988benzene} which, by breaking the molecular symmetry of benzene can cause the appearance of new spectral features.} The CO:phenol spectra does not contain any new features, but instead presents substantial alterations to features present in the undiluted phenol ice spectra. First, the OH stretch in the 3600-3000 \cmu range is significantly attenuated and two peaks at 3590 and 3575 \cmu are now present. CO has been previously observed to affect the OH feature belonging to H$_2$O in an analogous way. \citet{Bouwman07} observed the sharpening of the OH stretching bands in a water:CO ice mixtures and explained this feature by the breaking up of the hydrogen-bonding network between two water molecules in favor of the formation of new hydrogen bonds with the matrix molecules.
The appearance of sharp peaks in the 3500-3600 \cmu region has also been previously observed in dilute (1:1000) phenol:CO ices. \citet{gebicki1984interactions} identified the bands at 3590 and 3575 \cmu as representative of the formation of hydrogen bonding between phenol and the CO molecules.
Second, in the 1300-1150 \cmu range, the phenol ice shows a broad mode at 1230 \cmu having weak shoulders at 1260 and 1198 \cmu while in the phenol:CO mixture the intensities of the peaks at 1260 and 1198 \cmu appear significantly more prominent than the 1233 \cmu peak. The variation of the relative intensities of the bands in the  1300-1150 \cmu range is also a symptom of the formation of hydrogen bond with the CO solvation sphere \citep{gebicki1984interactions}. 

\begin{figure*}[thb!]
  \centering
  \includegraphics[width=\textwidth]{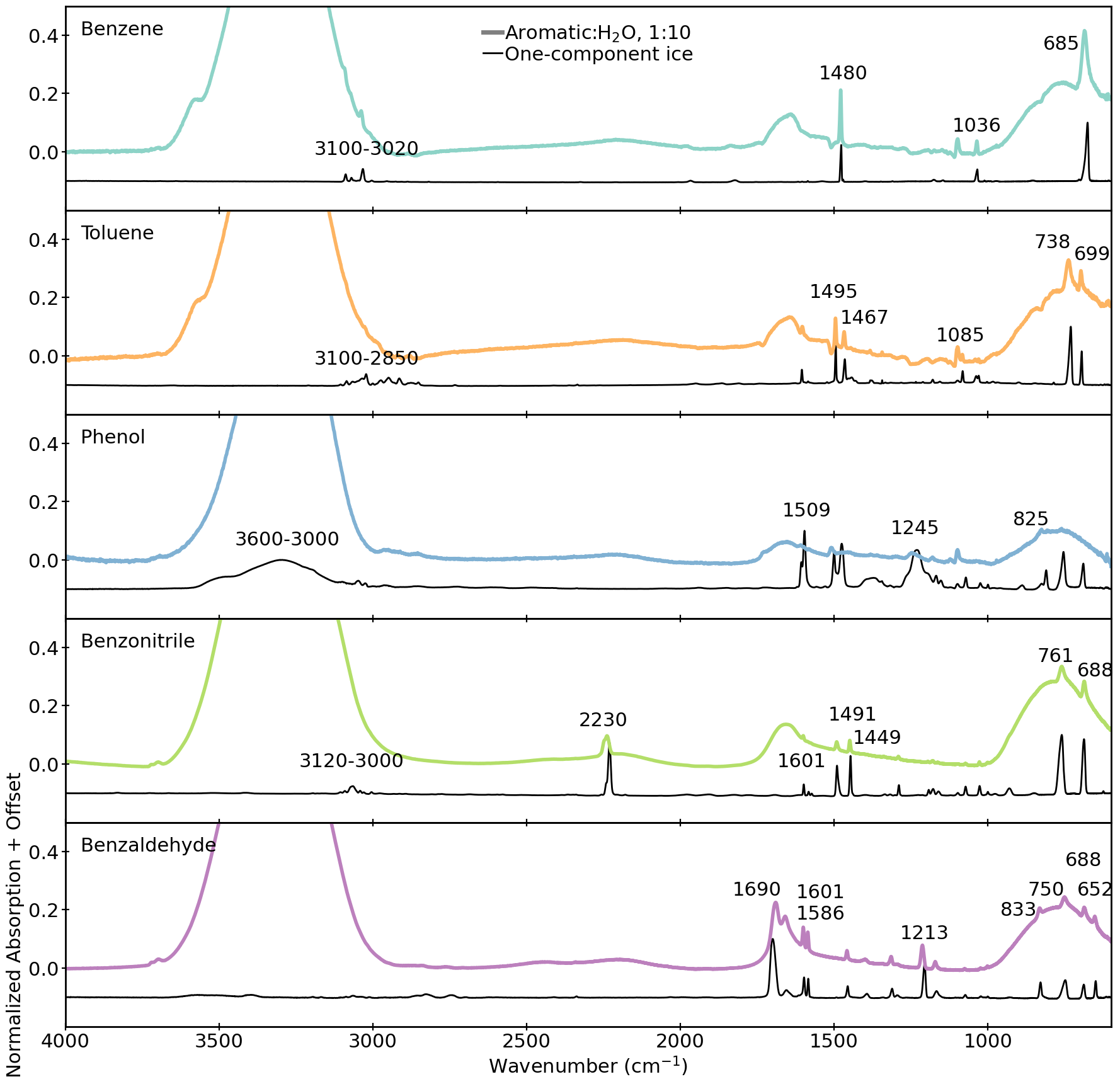}
 \caption{Normalized spectra of aromatic:H$_2$O 1:10 mixture ices condensed at 10 K.}
 \label{fig:IR_H2O_sub}
\end{figure*}

\subsubsection{H$_2$O ice matrix} \label{subsubh2o}
The spectra of monosubstituted benzene species in water are all characterized by weaker features and more ambiguous spectral baselines due to the strong and broad water features. However, the strongest IR features for all the molecules are still present. In the benzene:H$_2$O ice spectra, similarly to the benzene:CO mixture, we see the presence of a peak at 1100\cmu. Also in this case we speculate that the appearance of this feature is caused by the breaking of the symmetry of benzene due to the formation of hydrogen bonding between the $\pi$ system and the water \citep{Suzuki1992Sci...257..942S}.
Peak broadening when mixed with H$_2$O is common for all the five molecules and is particularly pronounced in the bands below 1000 \cmu. Band broadening in mixed ice is common and it is caused by the strong bonding of the molecules with the surrounding solvent (see Fig. \ref{fig:bro} in the Appendix \ref{appA} for an expansion on some key spectral features) \citep{tylli1986self,plokhotnichenko2001dimers}. Most of the aromatic peak positions are shifted by 1-12 \cmu to higher wavenumbers though this is more difficult to quantify compared to CO due to more ambiguous baseline subtraction.  (Table \ref{Table_big}). The one molecule that does not fit into this pattern is phenol, likely due to the similar hydrogen bonding capabilities between phenol and water which cause the formation of stronger solute-solvent interactions. Several bands are obscured by water in the spectra of the water:phenol mixture with only the strongest phenol bands remaining visible. (1503, 1235, 1072, and 825 \cmu in the undiluted ice). With the exception of the 825 \cmu band, which did not show any shift, all the other bands shifted significantly with the features in the mixture appearing at 1509, 1245, and 1100 \cmu respectively. 

\subsection{Spectral variation with ice temperature} \label{Sec:IR_VS_T}

During the ice TPDs we recorded IR spectra every 10 K, which are presented here for the undiluted aromatic ices spectra (Fig. \ref{fig:IR_H2O_T} and in Appendix \ref{appB} Fig. \ref{fig:IR_T} and \ref{fig:IR_CO_T} for the water and CO ice mixtures respectively). There are only minor changes with temperature in the water ice mixture. In the case of CO, the spectral evolution is close to identical to the undiluted ices due to the nearly complete CO sublimation around 30 K.

\begin{figure*}[thb!]
  \centering
  \includegraphics[width=\textwidth]{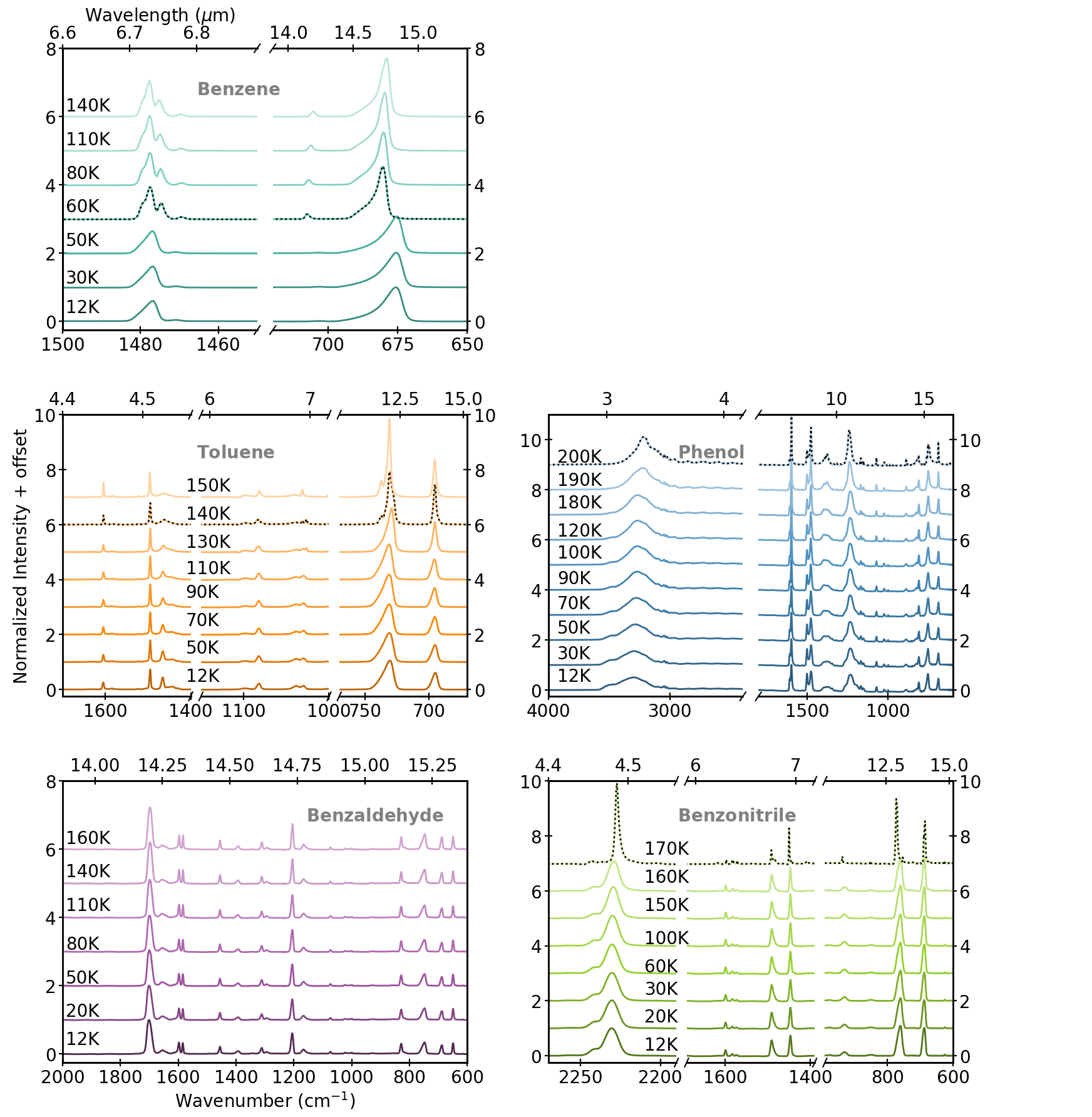}
 \caption{Evolution with temperature of the main IR features of each molecule undiluted aromatic ices IR spectra. The temperature at which a transition is observed is highlighted in dashed lines.}
 \label{fig:IR_T}
\end{figure*}

In the case of benzene ice we observe a variation in the spectral morphology between 50 and 60 K (Fig. \ref{fig:IR_T}). 
At this temperature range the feature at 676 \cmu, which is due to the out-of-plane  bending of the hydrogen atoms on the ring, shifts to 680 \cmu and an additional feature at 706 \cmu, which is due to ring deformation mode \citep{nna-mvodo2022ApJ...925..123N}, appears. A similar change can be observed in the same temperature range for the C-C stretch peak at 1477 \cmu where a new feature, likely arising from combination vibrational modes, sharpens at 1470 \cmu. These spectral variation are consistent with the ice-phase transition temperature of 55-60 K reported in previous studies for benzene \citep{mouzay2021experimental,ishii1996amorphous}. Our TPD experiments show that the crystallization is not yet complete by the time the desorption of the amorphous benzene is begins (120-140 K) and the fraction of the benzene that is still in its amorphous phase sublimates slightly before the crystalline ice (See \S\ref{sec:TPD} for additional details).  
For toluene the variation in the IR spectra as the temperature increases during the TPD experiment are not as dramatic as for benzene and they only occur right before desorption at 140 K. 

Differently from the other molecules studied, phenol does not present a well defined transition temperature but rather slowly rearranges from the amorphous to a more crystalline phase as the temperature increases. In particular we find that the intensity of the 3500 \cmu shoulder on the O-H band decreases with increasing temperature, while some fringes appear in the 2500-3100 \cmu range adjacent to the same OH band (Fig. \ref{fig:IR_T}), as the temperature increases. This phenomenon is possibly due to the formation of more organized trimers and multimers (2500-3100 \cmu range) substructures during crystallization  to the detriment of the monomeric state (3500 \cmu). This self association  phenomenon has been previously observed in matrix isolated phenol ices and it appears to shift spectral features to lower frequencies and increase peak intensity as the dilution decreases. \citep{tylli1986self,plokhotnichenko2001dimers}. We speculate that a behavior similar to matrix isolation experiments can be expected in undiluted ices when warming conditions are very slow and multiple nucleation points for the crystallization can form. In this scenario the phenol in the amorphous phase acts as a solvent for the more organized and less abundant nucleation sites, consequently causing a behavior that can be similar to the one observed in matrix isolated phenol conditions.

In the case of benzonitrile, desorption begins above 170 K. While the spectral profile stays quite consistent up until 160 K. We observe a sharp variation in the relative intensity of the peaks At 160 K compared to lower temperatures. At this temperature a peak broadening is observed across the spectra with the exception of the bands at 1163, 1181 and 2233 \cmu which become more narrow at the transition temperature.

Benzaldehyde desorption begins just below 180 K. The IR spectra does not show any significant variation during the warm up of the ice. The lack of a defined transition phase in the 10-160 K range, suggest that the formation of crystalline benzaldehyde, if occurring, might lie within the same range of temperature at which we observe benzaldehyde sublimation and consequently it cannot be observed in our experimental conditions.

\section{Desorption and binding energies of monosubstituted benzene molecules} \label{sec:TPD}

TPD profiles are useful to characterize the overall  desorption behaviors and binding energies of molecules. We ran three TPD experiments for each molecule (Table \ref{Table:exp}). In the next two subsections we discuss the desorption behavior of each molecule (\S\ref{sec:DES}), and then present an estimation of their binding energies (\S\ref{sec:BE}).

\begin{deluxetable}{c||ccc}
\tablecolumns{4}
\label{Table:exp}
\tablewidth{\textwidth}
\tablecaption{List of TPD experiments.}
\tablehead{
Molecule & \multicolumn{3}{c}{Ice coverage (ML) }}
\startdata
Benzene	      & 125	  & 275 & 436  \\
Toluene	      & 87    & 283   & 494  \\
Phenol        & 58	  & 64 & 389   \\
Benzonitrile  & 123 & 1106 & 1442 \\
Benzhaldehyde & 40    & 106 & 256  \\
\enddata
\vspace{-0.1cm}
\tablenotetext{}{All ices were formed at 10K. The uncertainty of the ice coverage is estimated to be within a factor of 2.} 
\end{deluxetable}

\subsection{TPD experiments}\label{sec:DES}
\begin{figure*}[thb!]
  \centering
  \includegraphics[width=\textwidth]{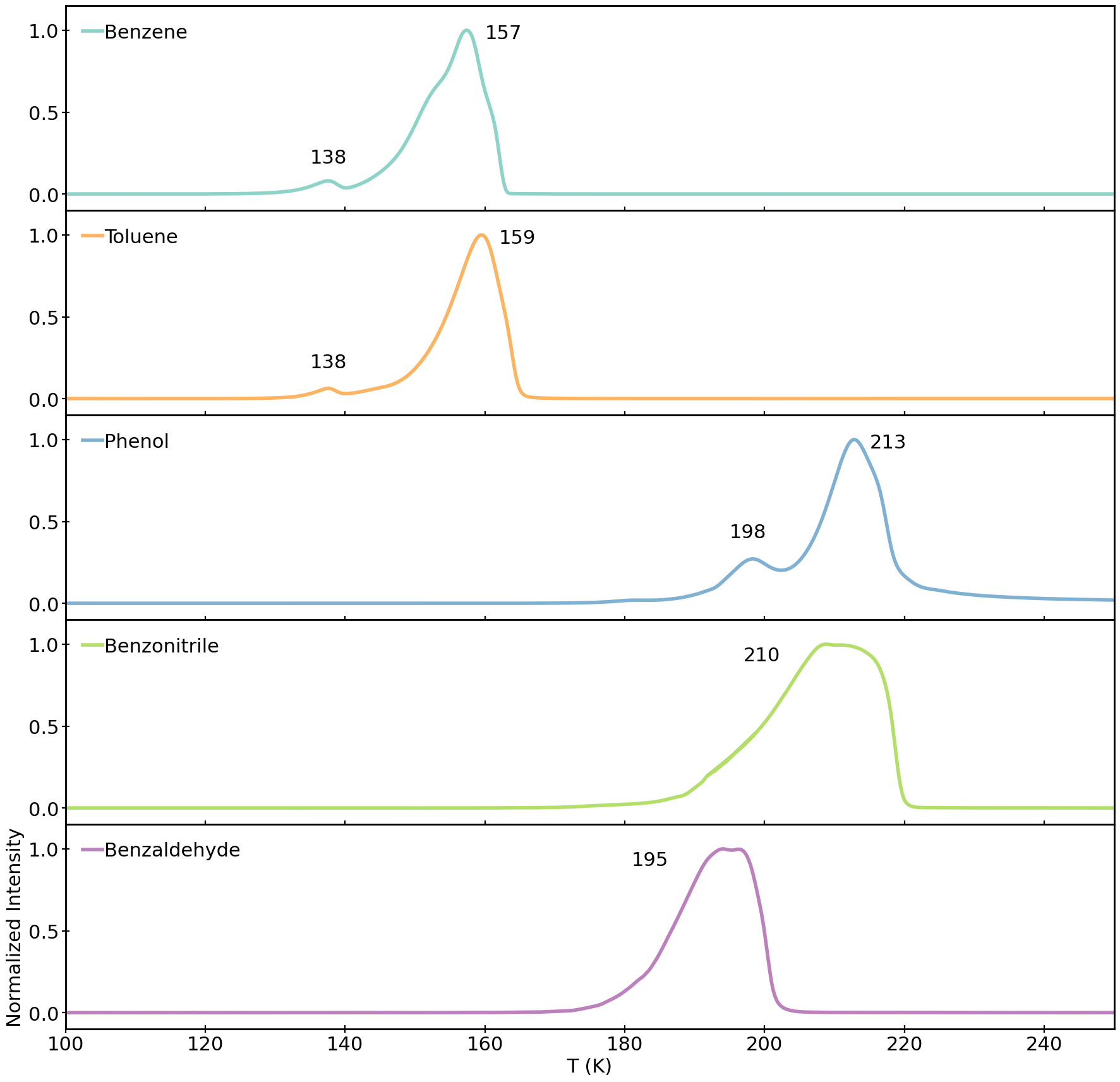}
 \caption{Normalized desorption profiles of the undiluted aromatic ices} during temperature programmed desorption (TPD) experiments
 \label{fig:tpd}
\end{figure*}

The TPD profiles of each molecule are shown in Fig. \ref{fig:tpd}. The desorption temperatures are generally quite high ($>$ 150 K) identifying these molecules as low-volatility From the TPD profiles it is also evident that most of the molecules undergo phase transitions during ice warm up which generally occurs just before the desorption onset, i.e. several of the curve show a double-peaked structure. A similar behavior is also observed in water ice \citep{fraser2001thermal}, indicating that the sublimation temperature of the amorphous phase of the ice falls very close to the range of temperatures necessary for the complete transition to crystalline ice. Consequently, the fraction of the ice that is still in its amorphous phase will desorb before having the chance of complete the phase transition.

For benzene we see an initial desorption at 138 K, followed by the main desorption peak at 155 K. This is consistent with the co-existence of two solid phases within the ice sample. Toluene has a similar desoption behavior as benzene, with the initial sublimation occurring at 138 K followed by a larger desorption peak at 160 K. Similarly to what observed in benzene ice it is likely that incomplete phase transition is responsible for the two desorption features of toluene.

Phenol also shows multiple desorption features. The main sublimation onset occurs between 205-210 K and its preceded by a shoulder around 195-200 K. A third weak desorption peak is present around 180 K, which is more evident in our thicker ices (Fig. \ref{fig:fit}). Phenol lacks a defined phase-transition temperature in the IR spectra (Fig. \ref{fig:IR_T}), but shows a slow change in the ice structure occurring at multiple nucleation points up to the onset of ice sublimation. This suggests that the phase transition may not uniformly occur in the entirety of the ice, therefore the presence of multiple desorption peaks is likely due to this blurred ice transition phase mechanisms which allows for multiple solid phases to coexist in the ice causing the appearance of several sublimation events.

Benzaldehyde presents a single desorption peak. This is consistent with the analysis of the IR spectra acquired during the TPD (\S\ref{Sec:IR_VS_T}) which does not show a phase transition at temperature below  the ice sublimation. However, the double peak structure of the TPD profile indicates that a phase transition is occurring in the same range as the benzaldehyde desorption temperature. For benzonitrile we see a variation in the IR spectra at 170 K just before the ice begins to sublimate in agreement with the TPD profile showing a small pre-peak just before ($\sim$10 K) the main sublimation onset at 190 K. 

Previous work shows that molecules with high molecular weight tend to have higher sublimation temperatures \citep{collings2004laboratory}. Additionally different functional groups will affect  both the long-range intermolecular interactions, such as in the formation of H-bond, as well as the the ring stacking capabilities of the aromatic rings in the short-range regime.

On the bases of molecular weight only we would expect the desorption temperature to follow the trend benzene (\mz  78) $<$ toluene (\mz  91) $<$ phenol (\mz  94) $<$ benzonitrile (\mz  103) $<$ benzaldehyde (\mz  106). While this is roughly true in our sample set, the nature of the functional groups present in the molecules also significantly affect the sublimation temperatures.

The effect of the substituent on the long-range molecular interactions is particularly evident in the case of phenol which has the highest sublimation temperature. The OH group on the phenol can form stronger hydrogen bonds compared to the other molecules studied. This adds to the stability of the phenol ice and shifts the sublimation temperature to values higher than what is expected from molecular weight considerations only.

The chemical nature of the functional groups also determine the intensity of short-range intermolecular interaction such as the stacking of aromatic rings with each other. The $\pi$-polarization-based view of stacking interactions \citep{cozzi1993dominance,hunter1990nature} suggests that the more a substituent is electron withdrawing (EW) the stronger the  $\pi$-$\pi$ interactions between two molecules in a sandwich configuration will be (i.e. interaction between two aromatic rings). Following this model we would expect the desorption temperature to follow the trend of benzene $<$ toluene $<$ phenol $<$ benzaldehyde $<$ benzonitrile. Once again phenol does not fit the expected trend and the energetic of the $\pi$-stacking configurations that are available to phenol compared to other molecules could help explain this behavior. \citet{zivkovic2019phenol} studied the potential energy curves for the parallel and antiparallel (with respect to the substituent) stacking of toluene and phenol rings. An example of the two extreme configurations (parallel and antiparallel) are shown in appendix \ref{confapp} Fig. \ref{conf} for phenol. They found that in toluene the antiparallel stacking is favorable and leads to a significantly stronger interaction compared to the parallel configuration. Phenol has no energetically driven preference in terms of stacking configuration and both the antiparallel and parallel configuration appear to be more stable than it is for toluene. When the ice is formed by quickly condensing molecules from the gas phase to 10 K, molecular rearrangement in the solid phase is energetically limited and the availability of several molecular configuration forming strong interactions, as is in the case of phenol, can results in an increased sublimation temperature of the ice.  

\subsection{Desorption Kinetics and Binding Energies} \label{sec:BE}

We next calculate the binding energies and attempt frequencies, which together define the desorption kinetics, of the five molecules from their TPD profiles.  As discussed in \S\ref{sec:DES}, benzene, toluene and phenol present with multiple desorption peaks corresponding to different ice morphologies. Ideally, it should be possible to derive a binding energy and a $\nu$ value for each desorption feature and, while we attempted to do so, we succeeded only in the case of benzene (see Table \ref{Table_BE}). In the case of toluene and phenol we speculate that due to a smaller temperature separation between the transition temperature and the desorption temperature of the amorphous phase, the TPD profile in that temperature range reflects multiple concomitant processes and consequently deviate from the desorption model used.

\begin{figure}[thb!]
  \centering
  \includegraphics[width=\columnwidth]{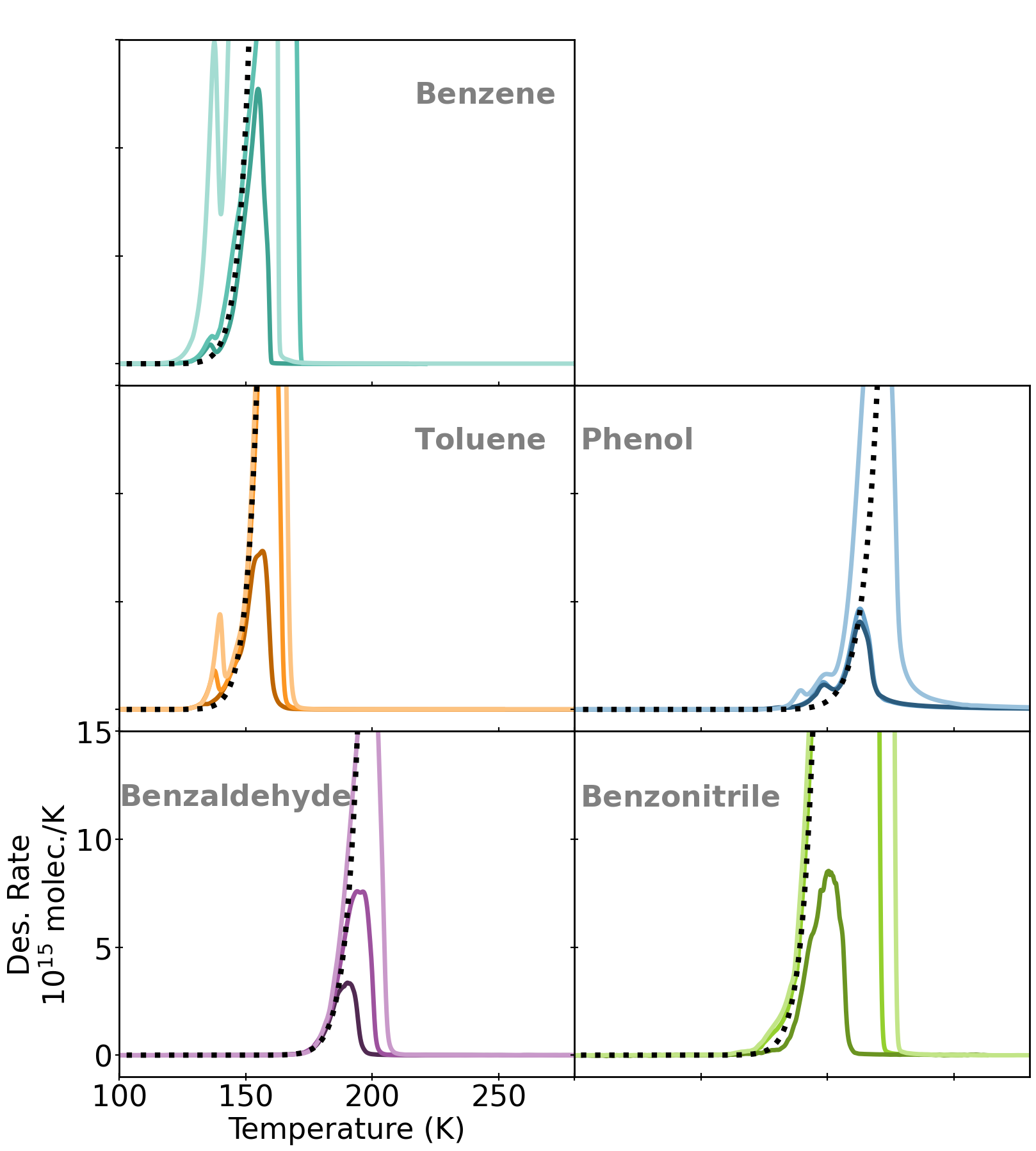}
 \caption{Experimental TPD curves and best fit curve in dashed lines.}
 \label{fig:fit}
\end{figure}

For each molecule we calculate the values of $\nu$ and of the binding energy corresponding to the main TPD desorption from the synchronous fit of three experimental data set using the Polyani-Wigner model (Eq.\ref{eqn1}). The resulting best fit curves are shown in Fig. \ref{fig:fit}. It is important to note that when multiple desorption peaks are present a range of temperature needs to be selected for the fit to avoid the inclusion of unwanted features. This has the consequence that in some cases the best fit curve might visually appear to be weighted toward a particular dataset.

For comparison, we also derive the binding energies associated with a value of $\nu$ calculated in the TST approximation ( Eq. \ref{eqn5}) and using the harmonic oscillator approximation (Eq. \ref{eqn2}). The attempt frequencies and binding energies values calculated using these three methods are reported in Table \ref{Table_BE}. 
The values of the theoretical $\nu$$_{TST}$ are always higher than the ones obtained from the empirical curve fit, and the predicted binding energies are therefore higher as well, by $\sim$ 1000-2000 K, but the difference decreases as the size of the molecule increases. By contrast, the $\nu$$_{har}$ values calculated using Eq. \ref{eqn2} are lower than the $\nu$$_{TPD}$ by a few orders of magnitude, resulting in lower binding energy estimates by up to 2000 K, with the largest differences for the most refractory molecules.
These trends are consistent with expectations that as the size of the molecules increases the traditional harmonic oscillator approach used to determine the pre-exponential factor of the Polyani-Wigner (See \S\ref{met:be}) equation should yield a less realistic description of the molecular behavior than the TST method. In other words, as the molecular complexity increases, including the molecular vibrational modes and geometry become more important to understand the desorption behavior.

As discussed in \S\ref{sec:met} the estimation of the attempt frequency of molecules is strongly dependent on the accuracy of the ice thickness. 
Since, we estimated the uncertainty on the thickness to be within a factor of two we tested the effects of the ice thickness on the goodness of the $\nu$ and BE fit by varying the value ice thickness used by +100$\%$ or -50$\%$ from the measured value. We found that these variation has a linear effect on the magnitude of the $\nu$ values, however, the same ice thickness variation only marginally affects the magnitude of the binding energy for all ices except benzene. The uncertainty in ice thickness and on the temperature calibration are the main source of error for the experimentally derived pre-exponential factor ($\nu$$_{TPD}$). The main source of error for the binding energy is also the absolute temperature calibration. 
In Table \ref{Table_BE} we report the uncertainty on the fit which includes all sources of experimental error.

The binding energy of benzene has recently been measured experimentally by \citet{salter_tpd_2018} and \citet{thrower_thermal_2009}. \citet{salter_tpd_2018} reports a binding energy value value for benzene of 5472.4 $\pm$ 216.5 K with and associated attempt frequency $\nu$ of $\sim$ 2.6x10$^{14}$ ML s$^{-1}$. \citet{thrower_thermal_2009} derived a binding energy value of 5785.1 $\pm$240 K in the multilayer regime, and a pre-exponential factor of the order of 10$^{14}$ ML s$^{-1}$ is calculated for benzene using a stochastic integration technique. Our values of 5220 K and 4.8x10$^{14}$ ML s$^{-1}$ (or 4.8x10$^{29}$ molecules cm$^{-2}$ s$^{-1}$, in the assumption that 1ML $\approx$ 10$^{15}$ molecules cm$^{-2}$) are consistent with both within uncertainties. For crystalline toluene, the binding energy of 5490$\pm$160 K that we derived from the synchronous fit of $\nu$ and the binding energy on the TPD (Eq. \ref{eqn2}), also compares well with the binding energy derived by \citet{salter_tpd_2018} of 5725$\pm$ 144.3 K. However the pre-exponential factor of 3.3x10$^{12}$ ML s$^{-1}$ derived by \citet{salter_tpd_2018} is a few order of magnitude lower than our value of 1.4x10$^{15}$ ML s$^{-1}$. 

We are not aware of any previous determination of the binding energies for phenol, benzonitrile and benzaldehyde, so a direct comparison with literature values cannot be done at this time. However the enthalpy of sublimation could be used as a proxy for binding energy values. Sublimation enthalpies have been reported for both benzene and phenol with values of 5340 \citep{de1980enthalpies} and 8251 K \citep{andon19601009} (44.4 and 68.6 kJ/mol) which well agree with our calculated binding energy values of 5220 and 8390 K respectively.

\begin{deluxetable*}{c||ccccccc}
\tablecolumns{8}
\tablecaption{Experimental zeroth order binding energy and pre-exponential factor.   
\label{Table_BE}}
\tablehead{Molecule	& T$_{max}$ & $\nu$$_{TPD}$ & BE(K)& $\nu$$_{TST}$ & BE$_{TST}$(K) & $\nu$$_{har}$ & BE$_{har}$(K) }
\startdata
Benzene$\textsuperscript{\textdagger}$        & A & (\textbf{6.9}$\pm$3.7)x10$^{13}$ &  \textbf{4780}$\pm$130   &  2.9x10$^{19}$  &   6754$\pm$30   &(9.7$\pm$0.1)x10$^{11}$ &   4380$\pm$138\\
                & C & (\textbf{4.8}$\pm$2.5)x10$^{14}$  &  \textbf{5220}$\pm$145   &  4.9x10$^{19}$  &   6925$\pm$32   &(1.1$\pm$0.1)x10$^{12}$ &   5000$\pm$157\\
Toluene$\textsuperscript{\textdagger}$        & C & (\textbf{1.4}$\pm$1.1)x10$^{15}$  &  \textbf{5490}$\pm$160    & 1.1x10$^{20}$  &   7195$\pm$85   &(9.1$\pm$0.1)x10$^{11}$ &   4570$\pm$151\\
Phenol          & C &\textbf{(5.4}$\pm$4.7)x10$^{16}$  &  \textbf{8390}$\pm$900  & 3.6x10$^{20}$  &   10180$\pm$120   &(1.0$\pm$0.2)x10$^{12}$ &   6365$\pm$209\\
Benzonitrile    & C & (\textbf{2.3}$\pm$2.1)x10$^{17}$  &  \textbf{7900}$\pm$240   & 4.8x10$^{20}$  &   9365$\pm$110   &(9.7$\pm$0.2)x10$^{11}$ &   5750$\pm$190\\
Benzaldehyde   & C & (\textbf{5.3}$\pm$4.4)x10$^{17}$  &  \textbf{8001}$\pm$193    & 4.0x10$^{20}$  &   9160$\pm$97   &(9.8$\pm$0.1)x10$^{11}$ &   5920$\pm$168\\
\enddata
\vspace{-0.1cm}
\tablenotetext{\textsuperscript{\textdagger}}{Literature values for benzene are BE=5472.4 $\pm$ 216.5 K, $\nu$ $\sim$ 2.6x10$^{14}$ ML s$^{-1}$  \citep{salter_tpd_2018} and BE=5785.1 $\pm$240 K, $\nu$ $\sim$  10$^{14}$ ML s$^{-1}$ \citep{thrower_thermal_2009}. The literature values for toluene are BE=5725$\pm$ 144.3 K.  $\nu$$\sim$3.3x10$^{12}$ ML s$^{-1}$ \citep{salter_tpd_2018}.}
\tablenotetext{*}{A= amorphous phase, C=crystalline phase, $\nu$$_{TPD}$= $\nu$ obtained from the TPD curve fit,$\nu$$_{TST}$= $\nu$ value calculated using the Transition State Theory,$\nu$$_{har}$= $\nu$ value estimated in the harmonic oscillator} approximation 
\tablenotetext{**}{The uncertainty reported include all source of experimental error including the uncertainty on the ice coverage and the absolute temperature calibration which constitute the main sources of error for $\nu$$_{TPD}$. The main source of uncertainty on the binding energy is the error on the temperature calibration. Please see text for more details.} 
\end{deluxetable*}

\section{Astrochemical implications and Conclusions} \label{sec:AI}
In this section we aim to illustrate the astrophysical implications of the laboratory experiments. First we estimate, using our experimental spectra and a publicly available JWST spectra from \citet{mclure2023NatAs...7..431M}, the upper limit for benzene towards a prestellar core. We next calculate the estimated snowline positions for the aromatic molecules in a simple disk model given our measured binding energies. 

\subsection{Observational Constraints}
 
A recent work by \citet{mclure2023NatAs...7..431M} identified a variety of molecules in two dense cores, including H$_2$O, CO$_2$, and CO, the main ice constituents, as well as rare isotopologues and more complex molecules such as CH$_3$OH and SO$_2$. Their spectra, includes the vibrational frequencies that are of interest for aromatics (600-1800 \cmu or 16.6-5.5 $\mu$m) possibly enabling the identification or the estimation of upper limits of aromatic molecules in the ice. In particular, the 5-8 $\mu$m range appears to be well suited for a comparison with experimental spectra of aromatic molecules as this portion of the spectra covers one of the main vibrational mode of benzene.  
We chose to ocus on benzene for this exercise, with the expectation of it being the most abundant of the five molecules studied.

\begin{figure}[thb!]
  \centering
  \includegraphics[width=\columnwidth]{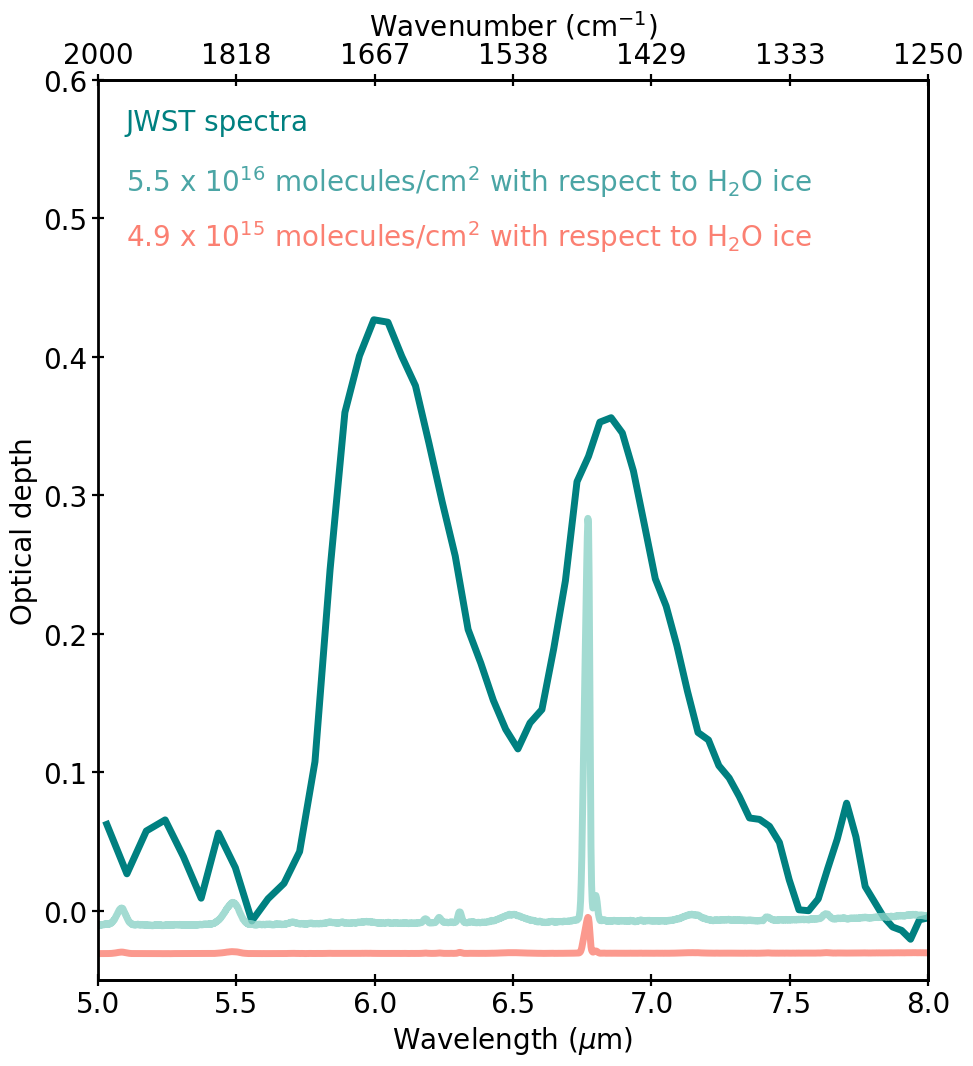}
 \caption{JWST spectra reproduced from \citet{mclure2023NatAs...7..431M} (teal) overlaid with the experimental benzene spectra having optical depth of the upper limit for detection (light blue) and optical depth estimated from the abundance of benzene in comet 67P \citep{aro67p2019A&A...630A..31S} (orange - offset).}
 \label{fig:JWST}
\end{figure}

In Fig. \ref{fig:JWST}, a portion of the spectra of a dense cloud at A$_v$$\sim$ 60 mag (NIR38) is reproduced from \citet{mclure2023NatAs...7..431M} and overplotted with the experimental benzene spectra. 
The continuum in the data from \citet{mclure2023NatAs...7..431M} has been modeled locally (5-12.1 $\mu$m) using a second order polynomial function. The model of the continuum was then subtracted from the original spectra which was also converted into optical depth.

In comparing the experimental benzene spectra with the JWST spectra, we find that the benzene feature at 1477 \cmu overlaps with a broad feature centered at about 6.8 \rm$\mu$ typically ascribed to methanol. We estimate a benzene upper limit by scaling the laboratory undiluted benzene spectra to close to the feature maximum (the blue trace in Fig. \ref{fig:JWST}), which corresponds to 5.5x10$^{16}$ molecules/cm$^2$ or a benzene upper limit of 1$\%$ compared to water ice. We can compare this limit to the expected abundance of benzene from cometary environments.
\citet{aro67p2019A&A...630A..31S} detected benzene in the coma of comet 67P with the ROSINA-DFMS and derived an abundance of 6.94x10$^{-6}$  -7.81x10$^{-4}$ with respect to water. We also show a spectrum of benzene scaled to the highest estimated comet abundance \citep{aro67p2019A&A...630A..31S} (Fig. \ref{fig:JWST}, orange trace), graphically showing the accuracy with which the 6.8 $\mu$m spectral feature needs to be modeled to be able to put interesting constraints on the benzene ice abundance.

It is interesting to note that at 6.72 $\mu$m there is a shoulder visible in the JWST spectra, this is very close to the 6.77 $\mu$m  band of benzene. In \citet{mclure2023NatAs...7..431M} this spectral feature is unidentified and attributed to unspecific COM functional groups. We speculate that could be due to a superposition of bands from benzene and other aromatic molecules, and suggest that careful modeling of JWST spectra in this region would be fruitful to constrain the overall ice aromatic abundance. Furthermore in lines of sight with more carbon-rich ices, this feature may become strong enough to identify with benzene if it is indeed its dominant carrier. 

\subsection{Aromatic molecules snowline predictions}
\begin{figure}[thb!]
  \centering
  \includegraphics[width=\columnwidth]{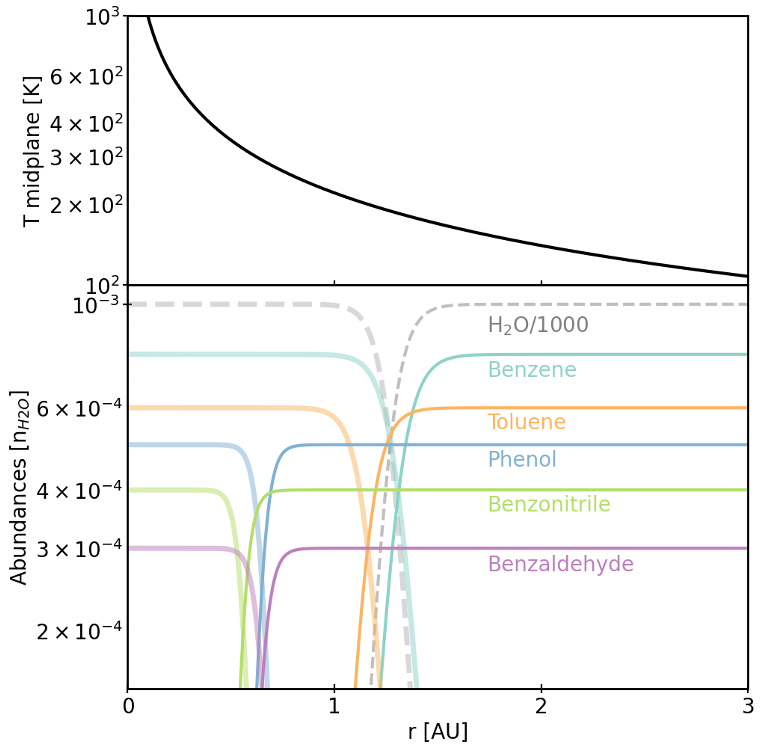}
 \caption{The top panel shows the temperature profile modelled in the midplane. The bottom panel shows the predicted snowline location of the five aromatic molecules.  }
 \label{fig:Snow}
\end{figure}
The desorption kinetics of molecules regulate the distribution of species in disks. The formation of condensation lines at different radius from the star is promoted by the radial variation in temperature observed in disks. This creates a gradient in the gas and ice distribution of molecules along the disk and consequently set the the composition of planets and planetesimals \citep{oberg2021astrochemistry}.

We estimated benzene and monosubstituted benzene snowline locations, using a simple model previously described in detail in \citet{oberg2019jupiter}. In brief, the model assumes the presence of a steady state between condensation and sublimation processes for each compound at any disk radius. Additionally, the midplane temperature is modelled as a radial power law. Under these assumptions, the model calculates the abundance of each molecule that is present in the gas/ice phase at each radius which is in turn dependent on the midplane temperature, the hydrogen density profile, and the total abundance of the molecule. In order to estimate the total abundance of aromatic molecules, we first used the median values reported for benzene in comet 67P \citep{aro67p2019A&A...630A..31S} and the water abundance from \citep{oberg2019jupiter}.
We then assumed the following lower abundances relative to H$_2$O for the different derivatives based on the the assumption that the parent molecule benzene is the most abundant - toluene, 6 x 10$^{-4}$; phenol, 5.0 x 10$^{-4}$; benzonitrile, 4.0 x 10$^{-4}$; benzaldehyde, 3.0 x 10$^{-4}$. However, we note that the snowline location does not depend on these assumptions as it is not dependent on the aromatic ice abundance but only on the desoprtion kinetics of the molecules.
Since all of these are rare species, we assume 1$^{st}$ order desorption with the same pre-exponential factor as derived in this paper but setting the surface coverage based on the aromatic ice abundance.

The predicted snowline positions obtained from this model are shown in Fig. \ref{fig:Snow}, where the top panel shows the calculated midplane temperature profile and the bottom panel shows the grain and gas traces for each of the molecules. Within the assumption of this model the condensation of small aromatics is estimated to occur at around 1 au. This result is consistent with the expectation that molecules having binding energies of above 5500 K should not be found in the gas phase at radius greater than 2 au where the water snowline is conventionally located \citep{oberg2019jupiter}. Consequently, the majority of the chemistry of small aromatic has to be expected to occur in the ice phase both in the ISM as well as across protoplanetary disks.

\section{Conclusions}
We surveyed a set of small aromatic molecules namely benzene, toluene, phenol, benzonitrile and benzaldehyde for their IR spectra as undiluted ices as well as when embedded in H$_2$O or CO ice matrices.
We further studied the desorption kinetics of each molecule estimating their corresponding binding energy values. Our experiments have provided:

\begin{enumerate}
    \item Detailed description of the IR spectra of aromatic molecules ices at 10 K as well as their spectral variability as a function of ice temperature. 
    \item The spectra of aromatic molecules in water and CO matrices showing that the presence of a CO matrix causes small variation in the aromatics spectra. Conversely in the presence of water the aromatic spectral features altered, and are sometimes only barely visible. 
    \item New experimental constraints on the desorption kinetics and binding energies of phenol, benzonitrile and benzaldehyde showing that they are low-volatility with binding energy values between 7900-8390 K. Most of the chemistry of these small aromatic molecules is therefore likely to occur in the ice phase in interstellar and circumstellar environments.
    \item Estimations of the abundance necessary for the detection of aromatics in dense clouds. The aromatic C=C stretching modes ($\sim$ 6.8 $\mu$m) could be detectable with JWST with scrupulous spectral modeling of the overlapping methanol band. 
    \item Estimation of the snowline position of small aromatic molecules in disks. The condensation of small aromatic molecules should be expected within 1.5 AU. 
\end{enumerate}

This work was supported by a grant from the Simons Foundation 686302, KÖ. and by an award from the Simons Foundation 321183FY19, KÖ.

\newpage
\bibliography{Characterization_of_monosubstituted_benzene}{}

\begin{thebibliography}{}
\expandafter\ifx\csname natexlab\endcsname\relax\def\natexlab#1{#1}\fi
\providecommand{\url}[1]{\href{#1}{#1}}
\providecommand{\dodoi}[1]{doi:~\href{http://doi.org/#1}{\nolinkurl{#1}}}
\providecommand{\doeprint}[1]{\href{http://ascl.net/#1}{\nolinkurl{http://ascl.net/#1}}}
\providecommand{\doarXiv}[1]{\href{https://arxiv.org/abs/#1}{\nolinkurl{https://arxiv.org/abs/#1}}}

\bibitem[{{Acharyya} {et~al.}(2007){Acharyya}, {Fuchs}, {Fraser}, {van Dishoeck}, \& {Linnartz}}]{Acharyya07}
{Acharyya}, K., {Fuchs}, G.~W., {Fraser}, H.~J., {van Dishoeck}, E.~F., \& {Linnartz}, H. 2007, \aap, 466, 1005, \dodoi{10.1051/0004-6361:20066272}

\bibitem[{{Allamandola} {et~al.}(1989){Allamandola}, {Tielens}, \& {Barker}}]{allamandola1989interstellar}
{Allamandola}, L.~J., {Tielens}, A.~G.~G.~M., \& {Barker}, J.~R. 1989, \apjs, 71, 733, \dodoi{10.1086/191396}

\bibitem[{Andon {et~al.}(1960)Andon, Biddiscombe, Cox, Handley, Harrop, Herington, \& Martin}]{andon19601009}
Andon, R., Biddiscombe, D., Cox, J., {et~al.} 1960, Journal of the Chemical Society (Resumed), 5246

\bibitem[{{Bergner} {et~al.}(2022){Bergner}, {Rajappan}, \& {{\"O}berg}}]{bergnerHCN2022ApJ...933..206B}
{Bergner}, J.~B., {Rajappan}, M., \& {{\"O}berg}, K.~I. 2022, \apj, 933, 206, \dodoi{10.3847/1538-4357/ac771e}

\bibitem[{{Bernstein} {et~al.}(1997){Bernstein}, {Sandford}, \& {Allamandola}}]{1997ApJ...476..932B}
{Bernstein}, M.~P., {Sandford}, S.~A., \& {Allamandola}, L.~J. 1997, \apj, 476, 932, \dodoi{10.1086/303651}

\bibitem[{{Bisschop} {et~al.}(2005){Bisschop}, {Fraser}, {Fuchs}, {{\"O}berg}, {Acharyya}, {van Broekhuizen}, {Schlemmer}, \& {van Dishoeck}}]{biss2005IAUS..231P.168B}
{Bisschop}, S.~E., {Fraser}, H.~J., {Fuchs}, G., {et~al.} 2005, in Astrochemistry: Recent Successes and Current Challenges, ed. D.~C. {Lis}, G.~A. {Blake}, \& E.~{Herbst}, Vol. 231, 168

\bibitem[{{Bouwman} {et~al.}(2010){Bouwman}, {Cuppen}, {Bakker}, {Allamandola}, \& {Linnartz}}]{bouwman2010A&A...511A..33B}
{Bouwman}, J., {Cuppen}, H.~M., {Bakker}, A., {Allamandola}, L.~J., \& {Linnartz}, H. 2010, \aap, 511, A33, \dodoi{10.1051/0004-6361/200913291}

\bibitem[{{Bouwman} {et~al.}(2007){Bouwman}, {Ludwig}, {Awad}, {{\"O}berg}, {Fuchs}, {van Dishoeck}, \& {Linnartz}}]{Bouwman07}
{Bouwman}, J., {Ludwig}, W., {Awad}, Z., {et~al.} 2007, \aap, 476, 995, \dodoi{10.1051/0004-6361:20078157}

\bibitem[{{Burkhardt} {et~al.}(2021){Burkhardt}, {Long Kelvin Lee}, {Bryan Changala}, {Shingledecker}, {Cooke}, {Loomis}, {Wei}, {Charnley}, {Herbst}, {McCarthy}, \& {McGuire}}]{burkhardt2021discovery}
{Burkhardt}, A.~M., {Long Kelvin Lee}, K., {Bryan Changala}, P., {et~al.} 2021, \apjl, 913, L18, \dodoi{10.3847/2041-8213/abfd3a}

\bibitem[{{Cannon} \& {Sutherland}(1951)}]{cannon1951infra}
{Cannon}, C.~G., \& {Sutherland}, G.~B.~B.~M. 1951, Spectrochimica Acta, 4, 373, \dodoi{10.1016/0371-1951(51)80008-0}

\bibitem[{{Cernicharo} {et~al.}(2001){Cernicharo}, {Heras}, {Tielens}, {Pardo}, {Herpin}, {Gu{\'e}lin}, \& {Waters}}]{cernicharo2001infrared}
{Cernicharo}, J., {Heras}, A.~M., {Tielens}, A.~G.~G.~M., {et~al.} 2001, \apjl, 546, L123, \dodoi{10.1086/318871}

\bibitem[{{Collings} {et~al.}(2004){Collings}, {Anderson}, {Chen}, {Dever}, {Viti}, {Williams}, \& {McCoustra}}]{collings2004laboratory}
{Collings}, M.~P., {Anderson}, M.~A., {Chen}, R., {et~al.} 2004, \mnras, 354, 1133, \dodoi{10.1111/j.1365-2966.2004.08272.x}

\bibitem[{{Cook} {et~al.}(2015){Cook}, {Ricca}, {Mattioda}, {Bouwman}, {Roser}, {Linnartz}, {Bregman}, \& {Allamandola}}]{cook2015photochemistry}
{Cook}, A.~M., {Ricca}, A., {Mattioda}, A.~L., {et~al.} 2015, \apj, 799, 14, \dodoi{10.1088/0004-637X/799/1/14}

\bibitem[{Cozzi {et~al.}(1993)Cozzi, Cinquini, Annuziata, \& Siegel}]{cozzi1993dominance}
Cozzi, F., Cinquini, M., Annuziata, R., \& Siegel, J.~S. 1993, Journal of the American Chemical Society, 115, 5330

\bibitem[{De~Kruif(1980)}]{de1980enthalpies}
De~Kruif, C. 1980, The Journal of Chemical Thermodynamics, 12, 243

\bibitem[{{Dubois} {et~al.}(2021){Dubois}, {Iraci}, {Barth}, {Salama}, {Vinatier}, \& {Sciamma-O'Brien}}]{Dubois2021PSJ.....2..121D}
{Dubois}, D., {Iraci}, L.~T., {Barth}, E.~L., {et~al.} 2021, \psj, 2, 121, \dodoi{10.3847/PSJ/ac06d5}

\bibitem[{Dunning~Jr(1989)}]{dunning1989gaussian}
Dunning~Jr, T.~H. 1989, The Journal of chemical physics, 90, 1007

\bibitem[{{Emery} {et~al.}(2023){Emery}, {Wong}, {Brunetto}, {Cook}, {Pinilla-Alonso}, {Stansberry}, {Holler}, {Grundy}, {Protopapa}, {Souza-Feliciano}, {Fern{\'a}ndez-Valenzuela}, {Lunine}, \& {Hines}}]{emery2023arXiv230915230E}
{Emery}, J.~P., {Wong}, I., {Brunetto}, R., {et~al.} 2023, arXiv e-prints, arXiv:2309.15230, \dodoi{10.48550/arXiv.2309.15230}

\bibitem[{{Fayolle} {et~al.}(2016){Fayolle}, {Balfe}, {Loomis}, {Bergner}, {Graninger}, {Rajappan}, \& {{\"O}berg}}]{Fayolle16}
{Fayolle}, E.~C., {Balfe}, J., {Loomis}, R., {et~al.} 2016, \apjl, 816, L28, \dodoi{10.3847/2041-8205/816/2/L28}

\bibitem[{{Ferrero} {et~al.}(2022){Ferrero}, {Grieco}, {Ibrahim Mohamed}, {Dulieu}, {Rimola}, {Ceccarelli}, {Nervi}, {Minissale}, \& {Ugliengo}}]{FerreroTST2022MNRAS.516.2586F}
{Ferrero}, S., {Grieco}, F., {Ibrahim Mohamed}, A.~S., {et~al.} 2022, \mnras, 516, 2586, \dodoi{10.1093/mnras/stac2358}

\bibitem[{{Fraser} {et~al.}(2001){Fraser}, {Collings}, {McCoustra}, \& {Williams}}]{fraser2001thermal}
{Fraser}, H.~J., {Collings}, M.~P., {McCoustra}, M. R.~S., \& {Williams}, D.~A. 2001, \mnras, 327, 1165, \dodoi{10.1046/j.1365-8711.2001.04835.x}

\bibitem[{Gebicki \& Krantz(1984)}]{gebicki1984interactions}
Gebicki, J., \& Krantz, A. 1984, Journal of the American Chemical Society, 106, 8093

\bibitem[{{Gerakines} \& {Hudson}(2015)}]{GerakinesCO2015ApJ...805L..20G}
{Gerakines}, P.~A., \& {Hudson}, R.~L. 2015, \apjl, 805, L20, \dodoi{10.1088/2041-8205/805/2/L20}

\bibitem[{{Gerakines} {et~al.}(2023){Gerakines}, {Materese}, \& {Hudson}}]{Gerakines2023MNRAS.522.3145G}
{Gerakines}, P.~A., {Materese}, C.~K., \& {Hudson}, R.~L. 2023, \mnras, 522, 3145, \dodoi{10.1093/mnras/stad1164}

\bibitem[{{Gerakines} {et~al.}(1995){Gerakines}, {Schutte}, {Greenberg}, \& {van Dishoeck}}]{bandstr1995A&A...296..810G}
{Gerakines}, P.~A., {Schutte}, W.~A., {Greenberg}, J.~M., \& {van Dishoeck}, E.~F. 1995, \aap, 296, 810, \dodoi{10.48550/arXiv.astro-ph/9409076}

\bibitem[{{Green} \& {Harrison}(1976)}]{green1976vibrational}
{Green}, J.~H.~S., \& {Harrison}, D.~J. 1976, Spectrochimica Acta Part A: Molecular Spectroscopy, 32, 1279, \dodoi{10.1016/0584-8539(76)80166-3}

\bibitem[{{Gudipati} \& {Allamandola}(2003)}]{gudipati2003ApJ...596L.195G}
{Gudipati}, M.~S., \& {Allamandola}, L.~J. 2003, \apjl, 596, L195, \dodoi{10.1086/379595}

\bibitem[{{Gudipati} \& {Allamandola}(2006)}]{gudipati2006ApJ...638..286G}
---. 2006, \apj, 638, 286, \dodoi{10.1086/498816}

\bibitem[{{He} {et~al.}(2016){He}, {Acharyya}, \& {Vidali}}]{he2016binding}
{He}, J., {Acharyya}, K., \& {Vidali}, G. 2016, \apj, 825, 89, \dodoi{10.3847/0004-637X/825/2/89}

\bibitem[{{Hudson} \& {Yarnall}(2022)}]{hudson2022infrared}
{Hudson}, R.~L., \& {Yarnall}, Y.~Y. 2022, \icarus, 377, 114899, \dodoi{10.1016/j.icarus.2022.114899}

\bibitem[{{Hudson} {et~al.}(2022){Hudson}, {Yarnall}, \& {Gerakines}}]{HudsonBenVP2022PSJ.....3..120H}
{Hudson}, R.~L., {Yarnall}, Y.~Y., \& {Gerakines}, P.~A. 2022, \psj, 3, 120, \dodoi{10.3847/PSJ/ac67a5}

\bibitem[{Hunter \& Sanders(1990)}]{hunter1990nature}
Hunter, C.~A., \& Sanders, J.~K. 1990, Journal of the American Chemical Society, 112, 5525

\bibitem[{Ishii {et~al.}(1996)Ishii, Nakayama, Yoshida, Usui, \& Koyama}]{ishii1996amorphous}
Ishii, K., Nakayama, H., Yoshida, T., Usui, H., \& Koyama, K. 1996, Bulletin of the Chemical Society of Japan, 69, 2831

\bibitem[{{Kaiser} {et~al.}(2015){Kaiser}, {Parker}, \& {Mebel}}]{kaiser2015ARPC...66...43K}
{Kaiser}, R.~I., {Parker}, D. S.~N., \& {Mebel}, A.~M. 2015, Annual Review of Physical Chemistry, 66, 43, \dodoi{10.1146/annurev-physchem-040214-121502}

\bibitem[{{K{\'o}sp{\'a}l} {et~al.}(2023){K{\'o}sp{\'a}l}, {{\'A}brah{\'a}m}, {Diehl}, {Banzatti}, {Bouwman}, {Chen}, {Cruz-S{\'a}enz de Miera}, {Green}, {Henning}, \& {Rab}}]{kospa2023ApJ...945L...7K}
{K{\'o}sp{\'a}l}, {\'A}., {{\'A}brah{\'a}m}, P., {Diehl}, L., {et~al.} 2023, \apjl, 945, L7, \dodoi{10.3847/2041-8213/acb58a}

\bibitem[{{Lampert} {et~al.}(1997){Lampert}, {Mikenda}, \& {Karpfen}}]{lampert1997molecular}
{Lampert}, H., {Mikenda}, W., \& {Karpfen}, A. 1997, Journal of Physical Chemistry A, 101, 2254, \dodoi{10.1021/jp962933g}

\bibitem[{{Leger} \& {Puget}(1984)}]{leger1984identification}
{Leger}, A., \& {Puget}, J.~L. 1984, \aap, 137, L5

\bibitem[{{Maksyutenko} {et~al.}(2022){Maksyutenko}, {Mart{\'\i}n-Dom{\'e}nech}, {Piacentino}, {{\"O}berg}, \& {Rajappan}}]{maksyutenko2022formation}
{Maksyutenko}, P., {Mart{\'\i}n-Dom{\'e}nech}, R., {Piacentino}, E.~L., {{\"O}berg}, K.~I., \& {Rajappan}, M. 2022, \apj, 940, 113, \dodoi{10.3847/1538-4357/ac94cb}

\bibitem[{{Mart{\'\i}n-Dom{\'e}nech} {et~al.}(2020){Mart{\'\i}n-Dom{\'e}nech}, {{\"O}berg}, \& {Rajappan}}]{martin2020formation}
{Mart{\'\i}n-Dom{\'e}nech}, R., {{\"O}berg}, K.~I., \& {Rajappan}, M. 2020, \apj, 894, 98, \dodoi{10.3847/1538-4357/ab84e8}

\bibitem[{{Materese} {et~al.}(2015){Materese}, {Nuevo}, \& {Sandford}}]{materese2015n}
{Materese}, C.~K., {Nuevo}, M., \& {Sandford}, S.~A. 2015, \apj, 800, 116, \dodoi{10.1088/0004-637X/800/2/116}

\bibitem[{{McClure} {et~al.}(2023){McClure}, {Rocha}, {Pontoppidan}, {Crouzet}, {Chu}, {Dartois}, {Lamberts}, {Noble}, {Pendleton}, {Perotti}, {Qasim}, {Rachid}, {Smith}, {Sun}, {Beck}, {Boogert}, {Brown}, {Caselli}, {Charnley}, {Cuppen}, {Dickinson}, {Drozdovskaya}, {Egami}, {Erkal}, {Fraser}, {Garrod}, {Harsono}, {Ioppolo}, {Jim{\'e}nez-Serra}, {Jin}, {J{\o}rgensen}, {Kristensen}, {Lis}, {McCoustra}, {McGuire}, {Melnick}, {{\~A}-berg}, {Palumbo}, {Shimonishi}, {Sturm}, {van Dishoeck}, \& {Linnartz}}]{mclure2023NatAs...7..431M}
{McClure}, M.~K., {Rocha}, W.~R.~M., {Pontoppidan}, K.~M., {et~al.} 2023, Nature Astronomy, 7, 431, \dodoi{10.1038/s41550-022-01875-w}

\bibitem[{{McGuire}(2018)}]{mcguire20182018}
{McGuire}, B.~A. 2018, \apjs, 239, 17, \dodoi{10.3847/1538-4365/aae5d2}

\bibitem[{{McGuire} {et~al.}(2021){McGuire}, {Loomis}, {Burkhardt}, {Lee}, {Shingledecker}, {Charnley}, {Cooke}, {Cordiner}, {Herbst}, {Kalenskii}, {Siebert}, {Willis}, {Xue}, {Remijan}, \& {McCarthy}}]{mcguire2021detection}
{McGuire}, B.~A., {Loomis}, R.~A., {Burkhardt}, A.~M., {et~al.} 2021, Science, 371, 1265, \dodoi{10.1126/science.abb7535}

\bibitem[{{McMurtry} {et~al.}(2016){McMurtry}, {Saito}, {Turner}, {Chakravarty}, \& {Kaiser}}]{mcmurtry2016formation}
{McMurtry}, B.~M., {Saito}, S. E.~J., {Turner}, A.~M., {Chakravarty}, H.~K., \& {Kaiser}, R.~I. 2016, \apj, 831, 174, \dodoi{10.3847/0004-637X/831/2/174}

\bibitem[{{Minissale} {et~al.}(2022){Minissale}, {Aikawa}, {Bergin}, {Bertin}, {Brown}, {Cazaux}, {Charnley}, {Coutens}, {Cuppen}, {Guzman}, {Linnartz}, {McCoustra}, {Rimola}, {Schrauwen}, {Toubin}, {Ugliengo}, {Watanabe}, {Wakelam}, \& {Dulieu}}]{MinissaleTST2022ESC.....6..597M}
{Minissale}, M., {Aikawa}, Y., {Bergin}, E., {et~al.} 2022, ACS Earth and Space Chemistry, 6, 597, \dodoi{10.1021/acsearthspacechem.1c00357}

\bibitem[{{Mooney}(1968)}]{kross1956infrared}
{Mooney}, E.~F. 1968, Spectrochimica Acta Part A: Molecular Spectroscopy, 24, 1999, \dodoi{10.1016/0584-8539(68)80260-0}

\bibitem[{{Mouzay} {et~al.}(2021){Mouzay}, {Couturier-Tamburelli}, {Pi{\'e}tri}, \& {Chiavassa}}]{mouzay2021experimental}
{Mouzay}, J., {Couturier-Tamburelli}, I., {Pi{\'e}tri}, N., \& {Chiavassa}, T. 2021, Journal of Geophysical Research (Planets), 126, e06566, \dodoi{10.1029/2020JE006566}

\bibitem[{{Nna-Mvondo} \& {Anderson}(2022)}]{nna-mvodo2022ApJ...925..123N}
{Nna-Mvondo}, D., \& {Anderson}, C.~M. 2022, \apj, 925, 123, \dodoi{10.3847/1538-4357/ac350c}

\bibitem[{{Noble} {et~al.}(2012){Noble}, {Theule}, {Mispelaer}, {Duvernay}, {Danger}, {Congiu}, {Dulieu}, \& {Chiavassa}}]{Noble12}
{Noble}, J.~A., {Theule}, P., {Mispelaer}, F., {et~al.} 2012, \aap, 543, A5, \dodoi{10.1051/0004-6361/201219437}

\bibitem[{{Nowak} {et~al.}(1988){Nowak}, {Menapace}, \& {Bernstein}}]{nowak1988benzene}
{Nowak}, R., {Menapace}, J.~A., \& {Bernstein}, E.~R. 1988, \jcp, 89, 1309, \dodoi{10.1063/1.455182}

\bibitem[{{Oberg}(2016)}]{oberg2016arXiv160903112O}
{Oberg}, K.~I. 2016, arXiv e-prints, arXiv:1609.03112, \dodoi{10.48550/arXiv.1609.03112}

\bibitem[{{{\"O}berg} \& {Bergin}(2021)}]{oberg2021astrochemistry}
{{\"O}berg}, K.~I., \& {Bergin}, E.~A. 2021, \physrep, 893, 1, \dodoi{10.1016/j.physrep.2020.09.004}

\bibitem[{{{\"O}berg} \& {Wordsworth}(2019)}]{oberg2019jupiter}
{{\"O}berg}, K.~I., \& {Wordsworth}, R. 2019, \aj, 158, 194, \dodoi{10.3847/1538-3881/ab46a8}

\bibitem[{{Plokhotnichenko} {et~al.}(2001){Plokhotnichenko}, {Radchenko}, {Blagoi}, \& {Karachevtsev}}]{plokhotnichenko2001dimers}
{Plokhotnichenko}, A.~M., {Radchenko}, E.~D., {Blagoi}, Y.~P., \& {Karachevtsev}, V.~A. 2001, Low Temperature Physics, 27, 666, \dodoi{10.1063/1.1399207}

\bibitem[{{Salter} {et~al.}(2021){Salter}, {Stubbing}, {Brigham}, \& {Brown}}]{salter2021using}
{Salter}, T.~L., {Stubbing}, J.~W., {Brigham}, L., \& {Brown}, W. 2021, Frontiers in Astronomy and Space Sciences, 8, 28, \dodoi{10.3389/fspas.2021.644277}

\bibitem[{{Salter} {et~al.}(2018){Salter}, {Stubbing}, {Brigham}, \& {Brown}}]{salter_tpd_2018}
{Salter}, T.~L., {Stubbing}, J.~W., {Brigham}, L., \& {Brown}, W.~A. 2018, \jcp, 149, 164705, \dodoi{10.1063/1.5051134}

\bibitem[{{Sankar} {et~al.}(2014){Sankar}, {Nowicka}, {Carter}, {Murphy}, {Knight}, {Bethell}, \& {Hutchings}}]{2014NatCo...5.3332S}
{Sankar}, M., {Nowicka}, E., {Carter}, E., {et~al.} 2014, Nature Communications, 5, 3332, \dodoi{10.1038/ncomms4332}

\bibitem[{{Schuhmann} {et~al.}(2019){Schuhmann}, {Altwegg}, {Balsiger}, {Berthelier}, {De Keyser}, {Fiethe}, {Fuselier}, {Gasc}, {Gombosi}, {H{\"a}nni}, {Rubin}, {Tzou}, \& {Wampfler}}]{aro67p2019A&A...630A..31S}
{Schuhmann}, M., {Altwegg}, K., {Balsiger}, H., {et~al.} 2019, \aap, 630, A31, \dodoi{10.1051/0004-6361/201834666}

\bibitem[{{Simon} {et~al.}(2023){Simon}, {Rajappan}, \& {{\"O}berg}}]{Simon2023ApJ...955....5S}
{Simon}, A., {Rajappan}, M., \& {{\"O}berg}, K.~I. 2023, \apj, 955, 5, \dodoi{10.3847/1538-4357/aceaf8}

\bibitem[{{Suzuki} {et~al.}(1992){Suzuki}, {Green}, {Bumgarner}, {Dasgupta}, {Goddard}, \& {Blake}}]{Suzuki1992Sci...257..942S}
{Suzuki}, S., {Green}, P.~G., {Bumgarner}, R.~E., {et~al.} 1992, Science, 257, 942, \dodoi{10.1126/science.257.5072.942}

\bibitem[{{Tabone} {et~al.}(2023){Tabone}, {Bettoni}, {van Dishoeck}, {Arabhavi}, {Grant}, {Gasman}, {Henning}, {Kamp}, {G{\"u}del}, {Lagage}, {Ray}, {Vandenbussche}, {Abergel}, {Absil}, {Argyriou}, {Barrado}, {Boccaletti}, {Bouwman}, {Caratti o Garatti}, {Geers}, {Glauser}, {Justannont}, {Lahuis}, {Mueller}, {Nehm{\'e}}, {Olofsson}, {Pantin}, {Scheithauer}, {Waelkens}, {Waters}, {Black}, {Christiaens}, {Guadarrama}, {Morales-Calder{\'o}n}, {Jang}, {Kanwar}, {Pawellek}, {Perotti}, {Perrin}, {Rodgers-Lee}, {Samland}, {Schreiber}, {Schwarz}, {Colina}, {{\"O}stlin}, \& {Wright}}]{tabone2023rich}
{Tabone}, B., {Bettoni}, G., {van Dishoeck}, E.~F., {et~al.} 2023, Nature Astronomy, 7, 805, \dodoi{10.1038/s41550-023-01965-3}

\bibitem[{{Terwisscha van Scheltinga} {et~al.}(2021){Terwisscha van Scheltinga}, {Marcandalli}, {McClure}, {Hogerheijde}, \& {Linnartz}}]{2021A&A...651A..95T}
{Terwisscha van Scheltinga}, J., {Marcandalli}, G., {McClure}, M.~K., {Hogerheijde}, M.~R., \& {Linnartz}, H. 2021, \aap, 651, A95, \dodoi{10.1051/0004-6361/202140723}

\bibitem[{{Thrower} {et~al.}(2009){Thrower}, {Collings}, {Rutten}, \& {McCoustra}}]{thrower_thermal_2009}
{Thrower}, J.~D., {Collings}, M.~P., {Rutten}, F.~J.~M., \& {McCoustra}, M.~R.~S. 2009, \jcp, 131, 244711, \dodoi{10.1063/1.3267634}

\bibitem[{{Tielens}(2013)}]{tielens2013molecular}
{Tielens}, A.~G.~G.~M. 2013, Reviews of Modern Physics, 85, 1021, \dodoi{10.1103/RevModPhys.85.1021}

\bibitem[{{Tielens} \& {Charnley}(1997)}]{tielens1997circumstellar}
{Tielens}, A.~G.~G.~M., \& {Charnley}, S.~B. 1997, Origins of Life and Evolution of the Biosphere, 27, 23, \dodoi{10.1023/A:1006513928588}

\bibitem[{{Tielens} \& {Hagen}(1982)}]{Tielens82}
{Tielens}, A.~G.~G.~M., \& {Hagen}, W. 1982, \aap, 114, 245

\bibitem[{{Tylli} \& {Konschin}(1986)}]{tylli1986self}
{Tylli}, H., \& {Konschin}, H. 1986, Journal of Molecular Structure, 142, 571, \dodoi{10.1016/0022-2860(86)85184-5}

\bibitem[{{Yang} {et~al.}(2022){Yang}, {Green}, {Pontoppidan}, {Bergner}, {Cleeves}, {Evans}, {Garrod}, {Jin}, {Kim}, {Kim}, {Lee}, {Sakai}, {Shingledecker}, {Shope}, {Tobin}, \& {van Dishoeck}}]{yang2022ApJ...941L..13Y}
{Yang}, Y.-L., {Green}, J.~D., {Pontoppidan}, K.~M., {et~al.} 2022, \apjl, 941, L13, \dodoi{10.3847/2041-8213/aca289}

\bibitem[{{Zacharia} {et~al.}(2004){Zacharia}, {Ulbricht}, \& {Hertel}}]{zacharia_interlayer_2004}
{Zacharia}, R., {Ulbricht}, H., \& {Hertel}, T. 2004, \prb, 69, 155406, \dodoi{10.1103/PhysRevB.69.155406}

\bibitem[{{Zeichner} {et~al.}(2023){Zeichner}, {Aponte}, {Bhattacharjee}, {Dong}, {Hofmann}, {Dworkin}, {Glavin}, {Elsila}, {Graham}, {Naraoka}, {Takano}, {Tachibana}, {Karp}, {Grice}, {Holman}, {Freeman}, {Yurimoto}, {Nakamura}, {Noguchi}, {Okazaki}, {Yabuta}, {Sakamoto}, {Yada}, {Nishimura}, {Nakato}, {Miyazaki}, {Yogata}, {Abe}, {Okada}, {Usui}, {Yoshikawa}, {Saiki}, {Tanaka}, {Terui}, {Nakazawa}, {Watanabe}, {Tsuda}, {Hamase}, {Fukushima}, {Aoki}, {Hashiguchi}, {Mita}, {Chikaraishi}, {Ohkouchi}, {Ogawa}, {Sakai}, {Parker}, {McLain}, {Orthous-Daunay}, {Vuitton}, {Wolters}, {Schmitt-Kopplin}, {Hertkorn}, {Thissen}, {Ruf}, {Isa}, {Oba}, {Koga}, {Yoshimura}, {Araoka}, {Sugahara}, {Furusho}, {Furukawa}, {Aoki}, {Kano}, {Nomura}, {Sasaki}, {Sato}, {Yoshikawa}, {Tanaka}, {Morita}, {Onose}, {Kabashima}, {Fujishima}, {Yamazaki}, {Kimura}, \& {Eiler}}]{Zeichner2023Sci...382.1411Z}
{Zeichner}, S.~S., {Aponte}, J.~C., {Bhattacharjee}, S., {et~al.} 2023, Science, 382, 1411, \dodoi{10.1126/science.adg6304}

\bibitem[{Zivkovic {et~al.}(2020)Zivkovic, Stankovic, Ninkovic, \& Zaric}]{zivkovic2019phenol}
Zivkovic, J.~M., Stankovic, I.~M., Ninkovic, D.~B., \& Zaric, S.~D. 2020, Crystal Growth \& Design, 20, 1025

\bibitem[{{Zwarich} {et~al.}(1971){Zwarich}, {Smolarek}, \& {Goodman}}]{zwarich1971assignment}
{Zwarich}, R., {Smolarek}, J., \& {Goodman}, L. 1971, Journal of Molecular Spectroscopy, 38, 336, \dodoi{10.1016/0022-2852(71)90118-4}

\end{thebibliography}
\bibliographystyle{aasjournal}

\newpage
\appendix{} \label{sec:appendix}

\section{Estimation of the IR band strengths}\label{appir}

Following \citet{1997ApJ...476..932B}, we estimate the band strengths by depositing a dilute and well-characterized mixture of an ice matrix species with known IR band strengths (CO) and each aromatic molecule. The method relies on the preparation of a gas-phase mixture in known proportion which is then condensed on the experimental substrate at 10 K. To minimize spectral and chemical interference we chose to use CO as the mixture solvent for mixtures prepared at a 1:10 ratio, aromatic:CO. In the assumption that the mixture ratio does not significantly change upon condensation and that the sticking coefficient of the mixture components have similar magnitude at the experimental temperature \citep{he2016binding,biss2005IAUS..231P.168B,Tielens82}, the CO ice coverage is measured using equation \ref{eqn0} along with the integrated absorbance of the CO band at 2139 \cmu. 

\begin{equation} \label{eqn0}
\rm N = \frac{2.3}{\rm A} \int Abs(\tilde\nu) d\tilde\nu
\end{equation}

where N is the column density of the molecule in molecules cm$^{-2}$. The coverage in monolayers (ML) can be obtained in the approximation that 1 ML=10$^{15}$ molecules cm$^{-2}$. $\int$Abs($\tilde\nu$) d$\tilde\nu$ is the area of the IR band in absorbance, and A is the band strength in cm molecule$^{-1}$.
Using the same equation in reverse, the band strength of the aromatic molecule can be then calculated assuming an ice thickness that is 1/10 of the one measured for CO. To estimate the ice coverage of CO we use the band strength value of 1.1x10$^{-17}$ \citep{bandstr1995A&A...296..810G}. 
All estimated band strength values are presented in Table \ref{table_A}. 
Benzene is the only molecule among the ones studied for which the band strengths are available in the literature. \citet{hudson2022infrared} measured amorphous and crystalline benzene ice band strengths with a 5-10$\%$ accuracy. They agree with our mixture-based estimates within 10$\%$ (35 $\%$ for the 676 \cmu band of benzene) suggesting that the mixture-based approach generally provides reasonable values. With only one comparison point, the uncertainty is, however, difficult to evaluate, and errors up to factor of two would not be shocking.
Based on the previous considerations we estimate that the aromatic molecule band strengths obtained with this methodology are accurate within a factor of two. These measurements should ideally be replaced by more precise band strength values measured using laser interferometry \citep{hudson2022infrared} which is currently not available on any of our experimental set-ups. In the meantime the reported band strengths will enable the evaluation of the ice coverage for the study of the desoprtion kinetics of the monosubstituted benzene molecules.

For toluene the most intense IR feature is at 720 \cmu, and this correspond to the highest A value in the spectra. The strength of toluene's IR bands does not vary as the temperature reaches the ice transition temperature. Given the chemical similarity with benzene we expect a smaller error for toluene than for some of the other molecules. The A values obtained for benzonitrile and benzaldehyde fall in the same range as the ones for benzene and toluene for similar ice features. 

The band strength values estimated for phenol are generally higher than the values estimated for the other aromatic molecules by a factors of 2-4 for analogous features. This is not expected based on existing gas-phase phenol band-strength values and we therefore speculate that our mixture based method may not work as well for phenol as for the other aromatics. Phenol is a solid at standard temperature and pressure (293.15 K and 1 atm) and its volatility is naturally very different than that of CO. This band strength estimate is hence highly uncertain, and our general error estimate of a factor of two may not fully capture the full uncertainty for this compound.

\begin{deluxetable}{c||cccc}
\tablecolumns{5}
\tablewidth{\columnwidth}
\label{table_A}
\tablecaption{Approximate band strengths values at 10 K (in CO)}
\tablehead{
Molecule & Band position & A/10$^{-18}$ 10 K & ref. Value at 10 K\tablenotemark{a}\\
 & (\cmu) &(cm molecule$^{-1})$& (cm molecule$^{-1})$}  
\startdata
Benzene	    & 676	& $\sim$10.8& 16.2  \\
            & 1036	& $\sim$1.8 & 1.94   \\
            & 1477	& $\sim$4.3 &  4.80  \\
 Toluene	   & 695	& $\sim$1.5 &-  \\
            & 720	& $\sim$6.0 &- \\
            & 1467	& $\sim$2.2 &- \\
            & 1495	& $\sim$3.0 &- \\
 Phenol	   & 693	& $\sim$6.3 &-\\
            & 755	& $\sim$16.3& - \\
            & 1478	& $\sim$10.7&  -\\
            & 1503	& $\sim$7.8 &- \\
            & 1598	& $\sim$13.1&  -\\
Benzonitrile& 689	& $\sim$2.2 &-\\
            & 763	& $\sim$4.6 &  -\\
            & 1447	& $\sim$1.4 & - \\
            & 1491	& $\sim$1.9 & - \\
            & 2230	& $\sim$3.4 & - \\
Benzaldehyde & 650	& $\sim$1.7 & -	\\
            & 690	& $\sim$1.6 &	-\\
            & 750	& $\sim$3.9 &-\\
            & 827	& $\sim$2.4 &	-\\
            & 1700	& $\sim$22.0&	\\
\enddata
\tablenotetext{a}{Reference band strengths are from \citet{hudson2022infrared}}
\tablenotetext{*}{The error associated with the band strength value is dependent on the error in the ice thickness for which we assume a factor of 2 uncertainty though, in reality, it is probably smaller for benzene and toluene, and perhaps higher for phenol.}
\end{deluxetable}

\newpage
\section{Spectral variation of the aromatic bands in the 1:10 aromatic:CO and aromatic:H$_2$O ice mixtures vs. undiluted ices ices}\label{appA}

\begin{figure}[thb!]
  \centering
  \includegraphics[width=0.7\textwidth]{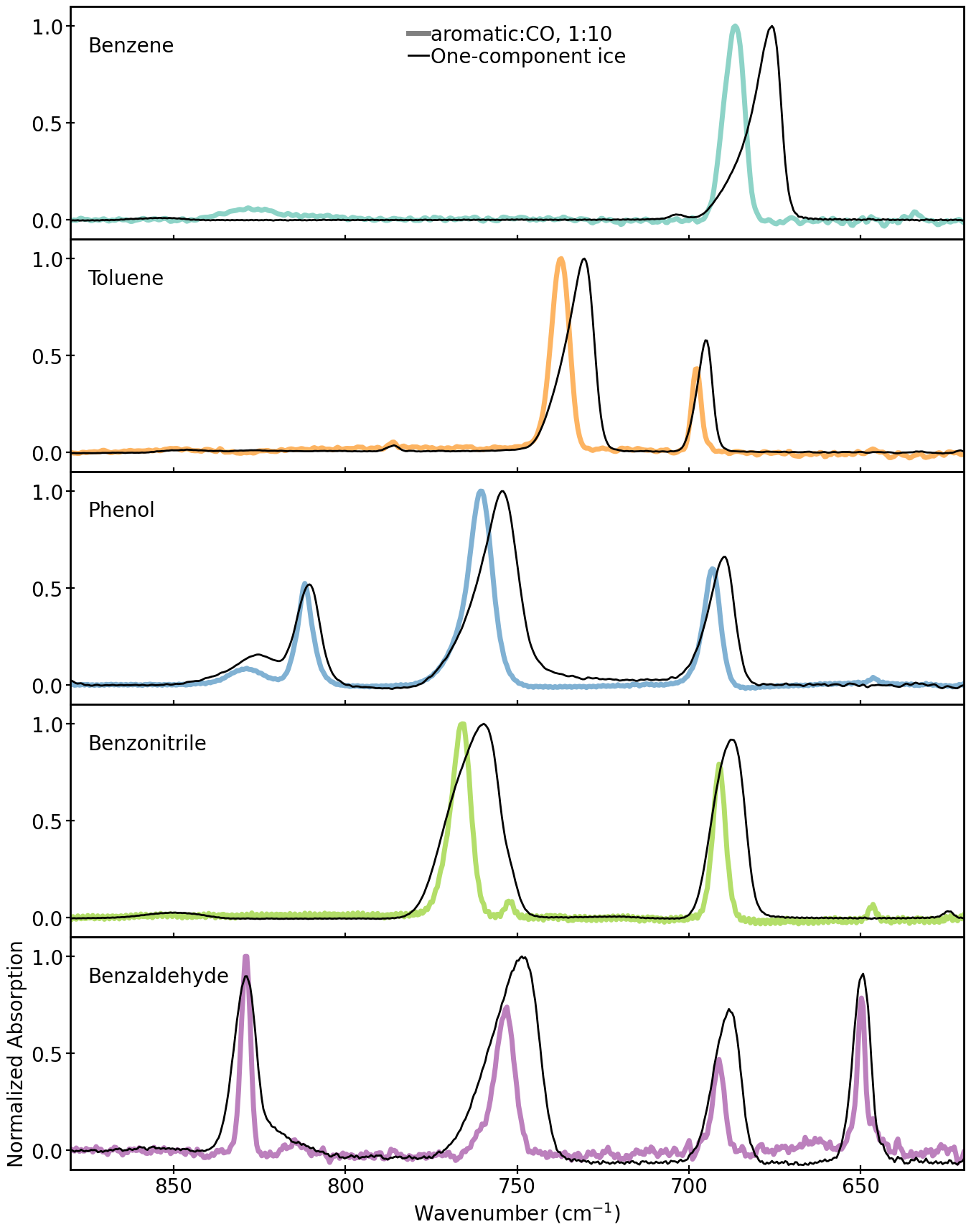}
 \caption{Spectra of undiluted aromatic ices (black) and aromatic:CO 1:10 ices. The bands in the 600-900 \cmu range appear narrower in the ice mixtures in comparison to the undiluted aromatic ices.}
 \label{fig:nar}
\end{figure}

\begin{figure*}[thb!]
  \centering
  \includegraphics[width=0.7\textwidth]{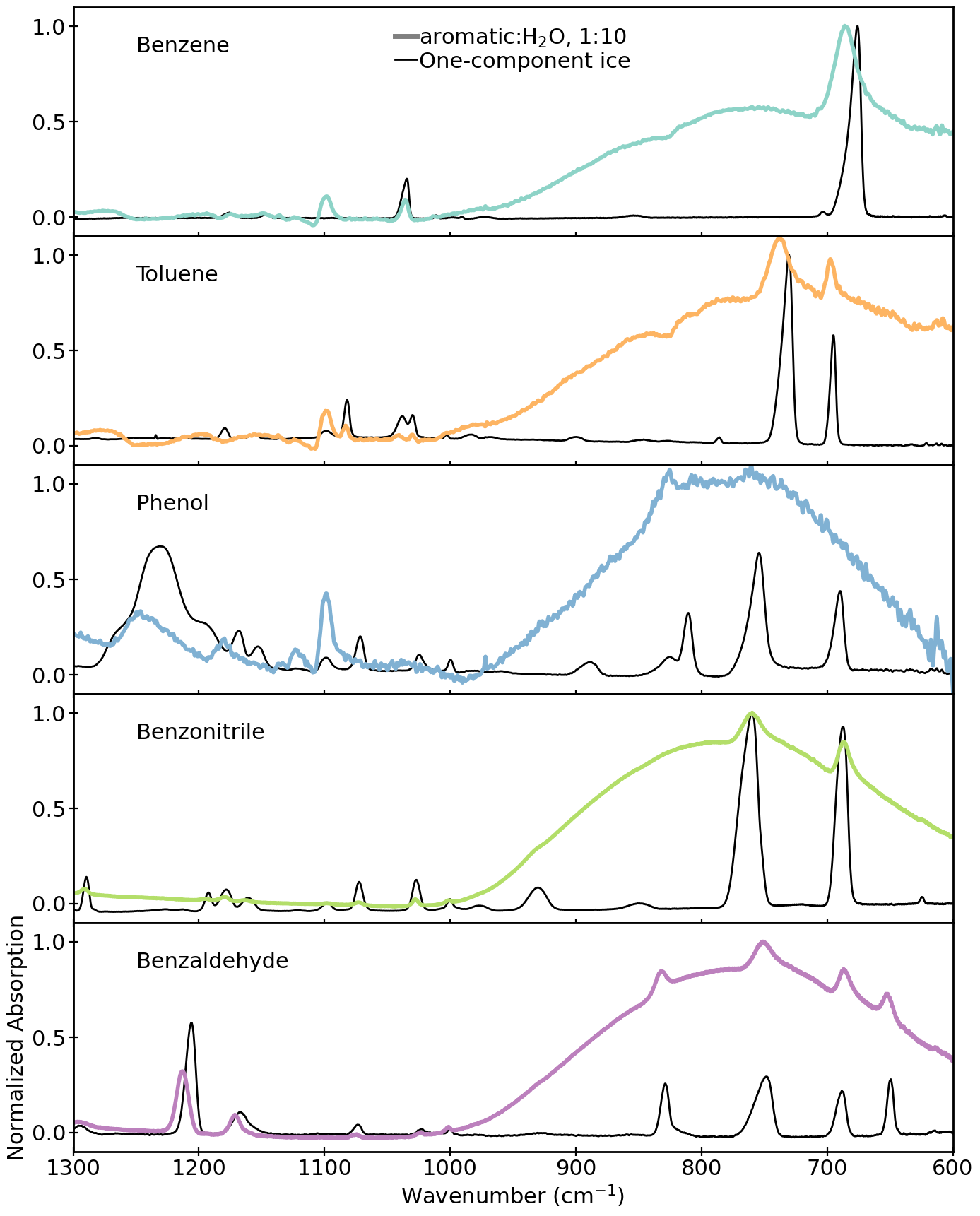}
 \caption{Spectra of undiluted benzene ice (grey) and aromatic:H$_2$O 1:10 ices. The aromatic molecules bands are broadened when mixed with water ice.}
 \label{fig:bro}
\end{figure*}

Figure \ref{fig:nar} show an expansion on the 900-600 \cmu spectral range for the 1:10 aromatic:CO ice mixtures. The shapes of the main aromatic bands are compared to highlight the band narrowing caused by the CO dilution.
In the case of aromatic:H$_2$O ice mixtures (Fig. \ref{fig:bro}) the aromatic molecules bands appear broader in the water mixure compared to the undiluted aromatic ices ices.

\newpage
\section{Spectral variation with temperature for the 1:10 aromatic:CO and aromatic:H$_2$O ice mixtures}\label{appB}

\begin{figure*}[thb!]
  \centering
  \includegraphics[width=0.7\textwidth]{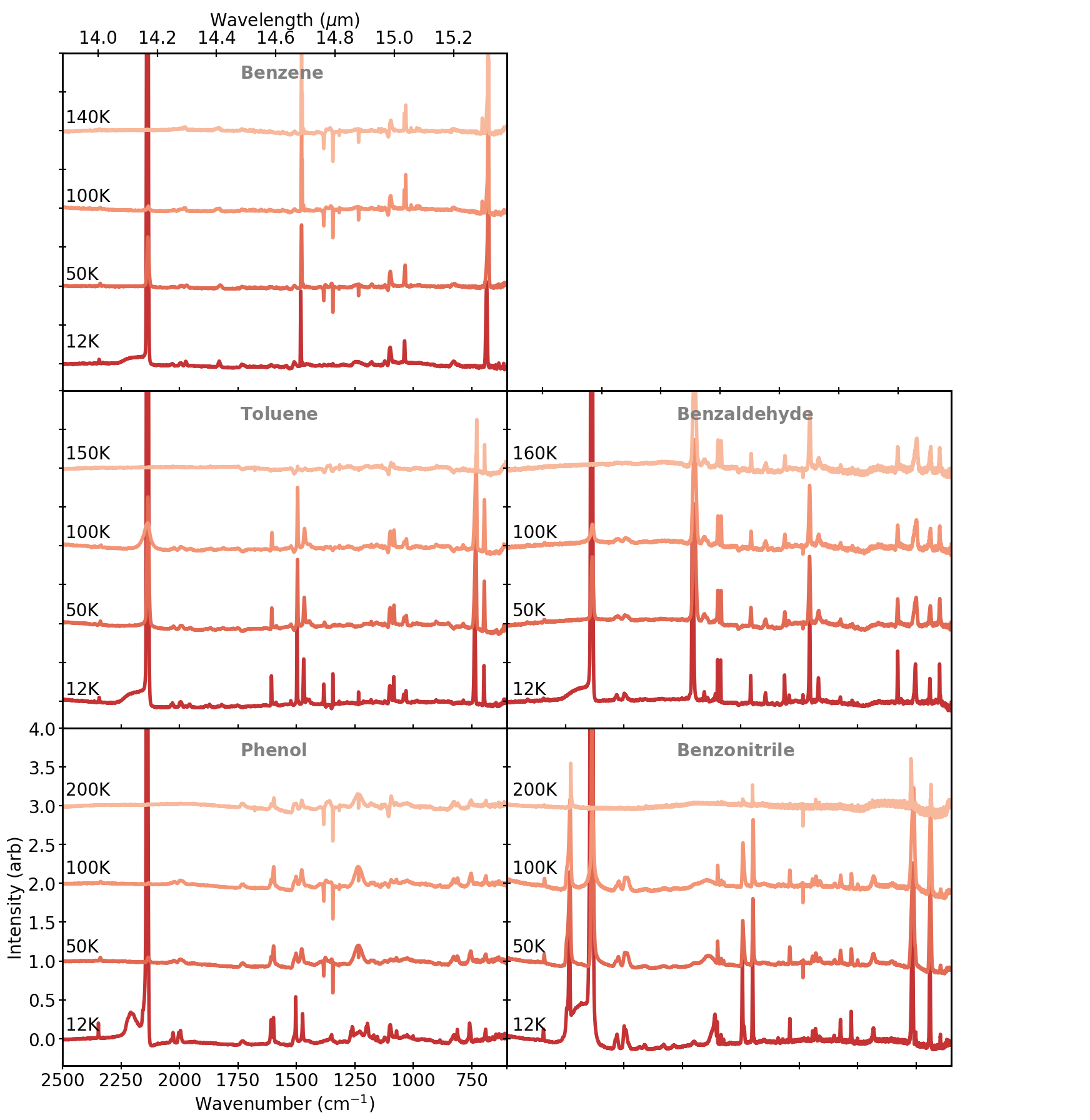}
 \caption{Spectral variation of an ice condensed from 1:10 mixture aromatic:CO as a function of temperature.}
 \label{fig:IR_CO_T}
\end{figure*}

\begin{figure*}[thb!]
  \centering
  \includegraphics[width=0.7\textwidth]{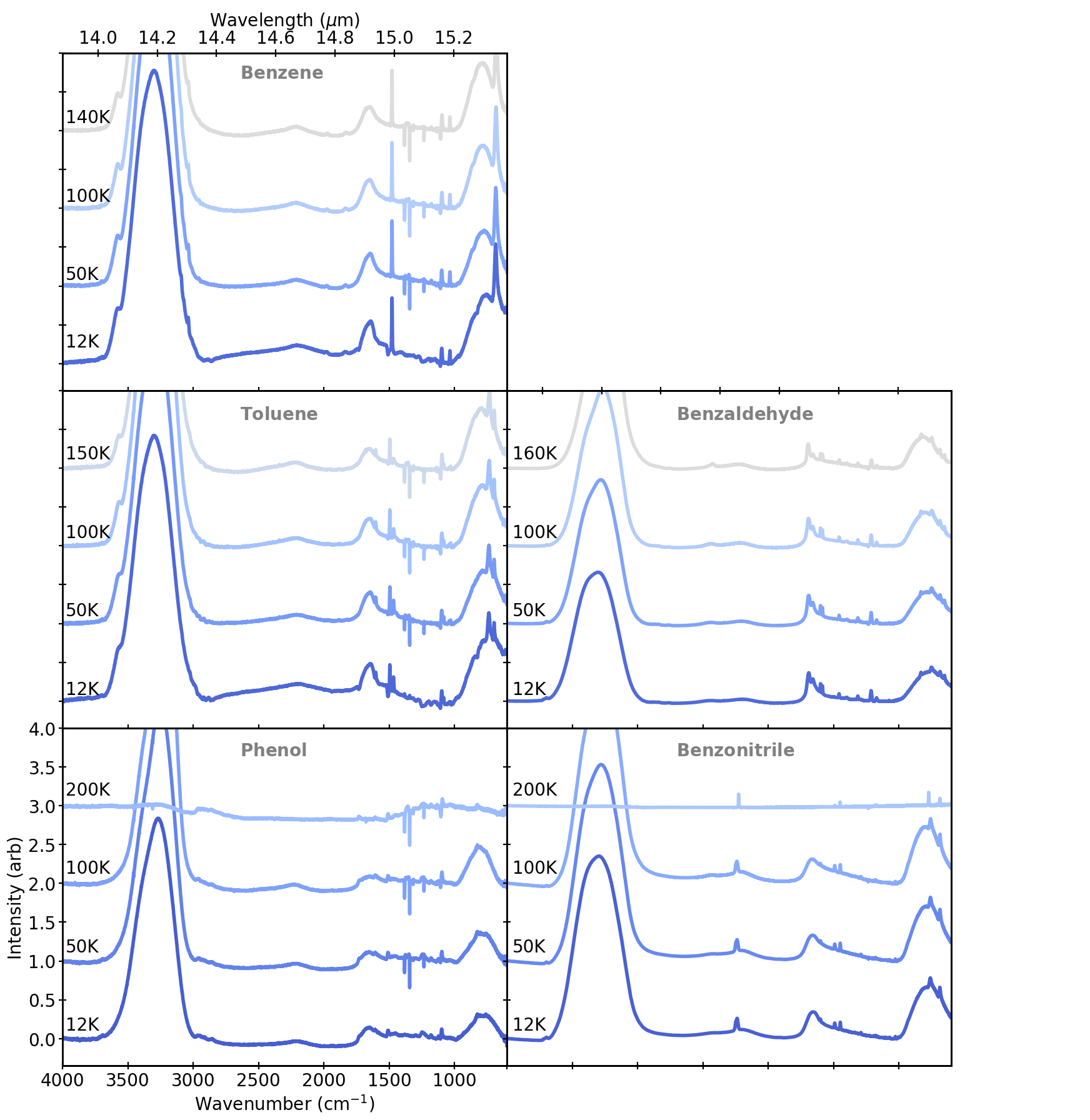}
 \caption{Spectral variation of an ice condensed from 1:10 mixture aromatic:H$_2$O as a function of temperature.}
 \label{fig:IR_H2O_T}
\end{figure*}

In the figures below we show the spectra of the 1:10 mixed ices (aromatic:CO Fig. \ref{fig:IR_CO_T}, and aromatic:H$_2$O  Fig. \ref{fig:IR_H2O_T}) at different temperatures during the TPD experiments. In both the H$_2$O and the CO sets of spectra we see the formation of downward pointing spectral artifacts in the 1250 to 1400 \cmu  region for temperature higher than 50K. Spectral variations with temperature in the CO mixtures are similar to what observed in the undiluted ices with the addition of the disappearance of the CO fundamental band at 2139 \cmu beginning at about 40K. In the aromatic water mixtures the aromatic features are muted and variations of the band shape with temperature are not pronounced.

\newpage
\section{Full range TPD curves}\label{appc}

\begin{figure*}[thb!]
  \centering
  \includegraphics[width=0.7\textwidth]{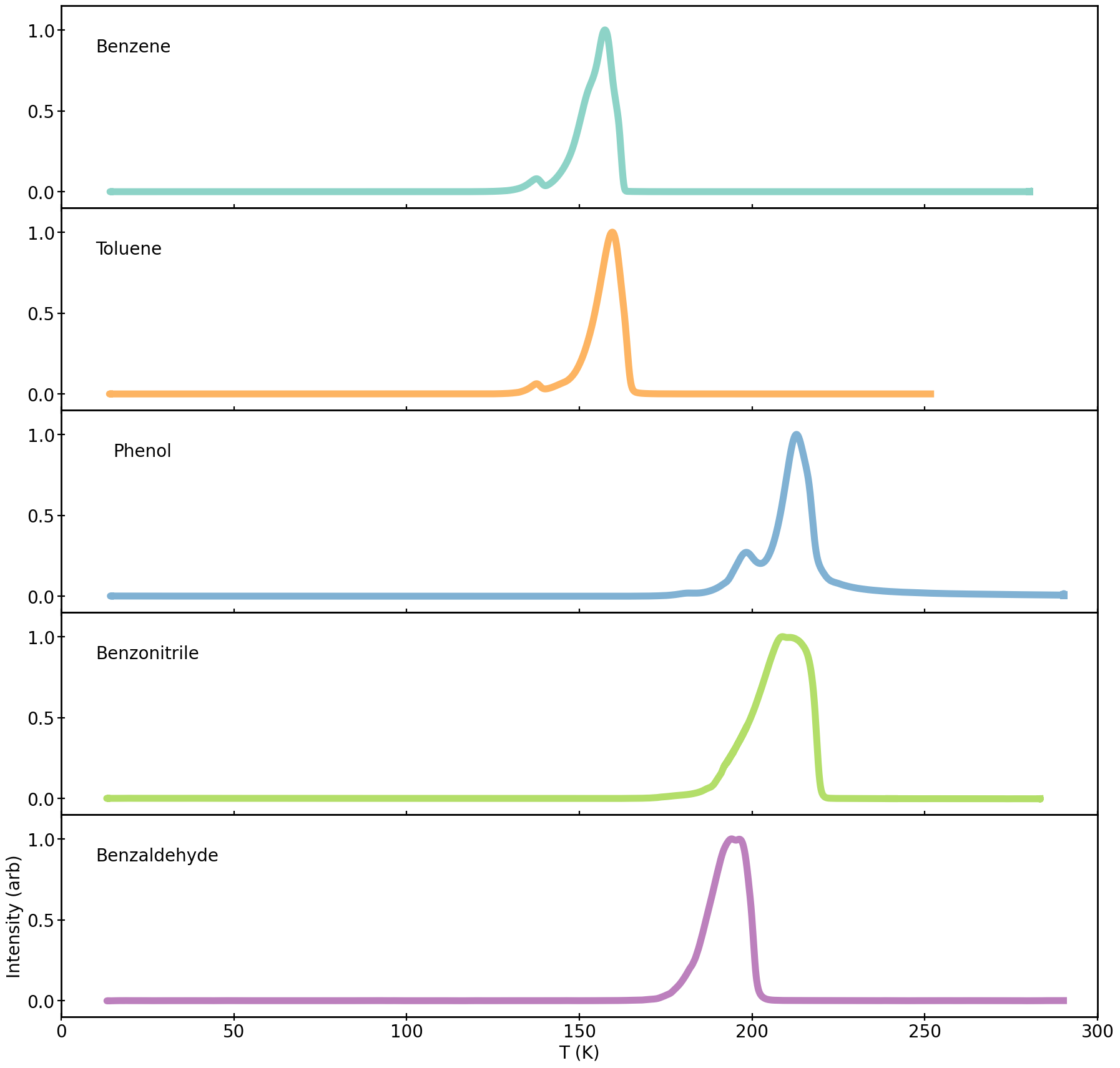}
 \caption{TPD profiles of each of the aromatic molecule along the full range of temperatures.}
 \label{fig:ms_app}
\end{figure*}

In Fig. \ref{fig:tpd} we show the TPD curves on reduced axes to allow for a more detailed visualization of the desorption peaks. For the sake of transparency we show the full range of the TPD curves in Fig. \ref{fig:ms_app}. No desorption peaks were observed below T=100 K.

\newpage
\section{Notes on the Benzonitrile TPD profile }\label{appd}

\begin{figure*}[thb!]
  \centering
  \includegraphics[width=0.5\textwidth]{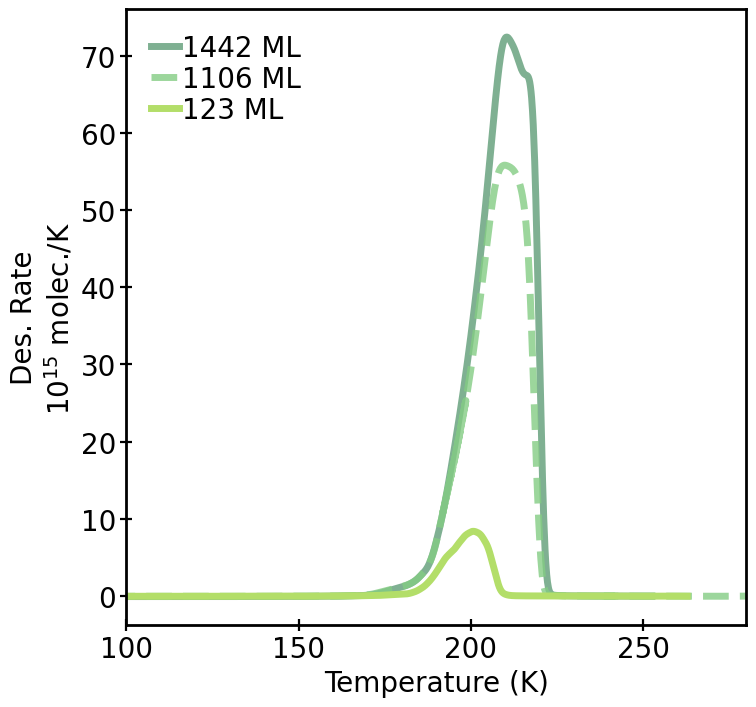}
 \caption{TPD profiles of the three benzonitrile experiments plotted on full axis. The dotted line highlight the experiment presented in figure \ref{fig:tpd}}
 \label{fig:ms_benit_app}
\end{figure*}

In figure \ref{fig:tpd} the desorption profile of the benzonitrile presents with a flat top which could be misread as a sign of saturation of the mass spectrometer. In figure \ref{fig:ms_benit_app} we show the variation of the feature with ice coverage which demonstrates that higher coverage ices still provide a distinct mass spectrometric signal. This hints to the existence of multiple desorption processes, likely due to different solid-phases of benzonitrile, overlapping in the same temperature range.

\newpage
\section{Calculation of the principal moment of inertia of the aromatic molecules}\label{inertia}

The transition state model used to estimate the pre-exponential factor computationally (See \S\ref{met:be}) necessitate of values of the principal moment of inertia ($I_{\rm x}I_{\rm y}$ and $I_{\rm z}$) and of the symmetry factor ($\sigma$) of each of the molecules. We choose to determine this values computationally by running electronic structure calculation using Gaussian 16 \citep{dunning1989gaussian}. For each molecule we run geometry optimization and frequency calculations at the B3LYP/aug-cc-pVDZ level of theory. The $I_{\rm x}$, $I_{\rm y}$, $I_{\rm z}$ and the $\sigma$ values are calculated as part of the vibrational frequency calculations. 
The values used in this work are reported in Table \ref{table_I}.

\begin{deluxetable}{c||cccc}
\tablecolumns{5}
\tablewidth{0.5\textwidth}
\label{table_I}
\tablecaption{Values of the principal moment of inertia and symmetry factors for the aromatic molecules. Values are calculated using Gaussian 16 at the B3LYP/aug-cc-pVDZ level of theory.}
\tablehead{
Molecule & $I_{\rm x}$ & $I_{\rm y}$ & $I_{\rm z}$ & $\sigma$\\
 & (amu a$_0$$^2$) & (amu a$_0$$^2$)&  (amu a$_0$$^2$) &   }
\startdata
Benzene	     & 318.7	& 318.7  & 637.4  & 1 \\
Toluene	     & 327.3	& 721.4  & 1037.4 & 1 \\
Phenol	     & 343.7	& 695.3  & 1039.0 & 1 \\
Benzonitrile & 320.0	& 1174.2 & 1494.6 & 1 \\
Benzaldehyde & 345.9	& 1162.7 & 1508.6 & 1 \\
\enddata
\end{deluxetable}

\newpage
\section{Parallel and antiparallel configuration of monosubstituted aromatic molecules} \label{confapp}

\begin{figure}[thb!]
  \centering
  \includegraphics[width=0.5\textwidth]{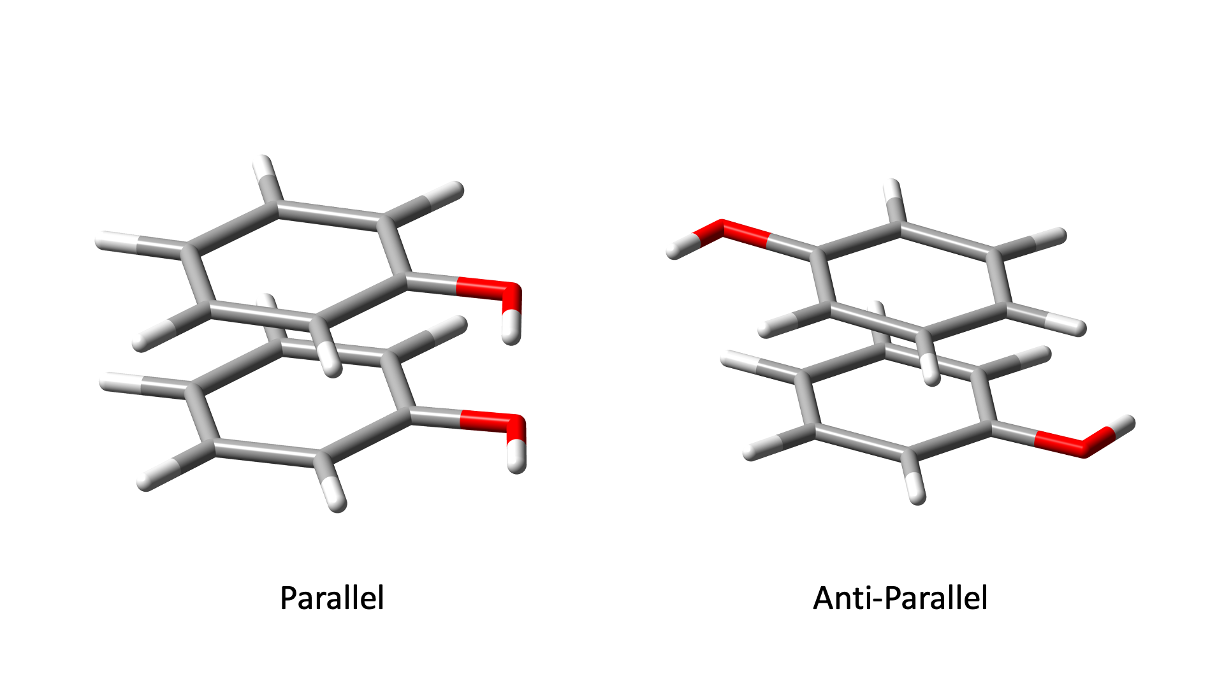}
 \caption{Visual representation of the parallel and antiparallel configuration for phenol.}
 \label{conf}
\end{figure}

\citet{zivkovic2019phenol} studied computationally the energetic of the interaction of toluene and phenol molecules in solid phase. They found that toluene has a preference toward the anti-parallel arrangements while phenol shows no preferential configuration. In Fig.\ref{conf} we show an example for phenol of the two most extreme configurations.

\end{document}